\documentclass[journal,draftclsnofoot,onecolumn,oneside,12pt]{IEEEtran}

\usepackage[T1]{fontenc}
\usepackage{amsmath} 
\usepackage{amssymb}
\usepackage{algorithmic}
\usepackage{cite}
\usepackage{dsfont} 
\ifCLASSINFOpdf
   \usepackage[pdftex]{graphicx}
\else
  \usepackage[dvips]{graphicx}
\fi

\newtheorem{property}{Property}
\newtheorem{theorem}{Theorem}
\newtheorem{proposition}{Proposition}
\newtheorem{claim}{Claim}
\newtheorem{definition}{Definition}

\DeclareMathOperator*{\argmin}{argmin}
\DeclareMathOperator*{\argmax}{argmax}
\DeclareMathOperator{\sgn}{sgn}

\begin{document}
\title{Maximum Signal Minus Interference to Noise Ratio Multiuser Receive
Beamforming}

\author{Majid Bavand, \IEEEmembership{Student Member, IEEE,} Steven D. Blostein,
\IEEEmembership{Senior Member, IEEE}
\thanks{Part of this work was presented at 27th Biennial Symposium on
Communications (QBSC), ON, June 2014, and was the runner-up for the best student paper award.

The authors are with the Department
of Electrical and Computer Engineering, Queen's University, Ontario,
Canada, K7L 3N6 (e-mail: \{m.bavand, steven.blostein\}@queensu.ca).}
}
\maketitle

\begin{abstract}
Motivated by massive deployment of low data rate Internet of things (IoT)
and ehealth devices with requirement for highly reliable communications,
this paper proposes receive
beamforming techniques for the
uplink of a single-input multiple-output (SIMO)
multiple access channel (MAC), based on a
per-user probability of error metric and one-dimensional signalling.
Although beamforming by directly minimizing probability of error (MPE)
has potential advantages over
classical beamforming methods such as zero-forcing and minimum
mean square error beamforming, MPE beamforming results
in a non-convex and a highly nonlinear optimization problem.
In this paper, by adding a set of modulation-based constraints,
the MPE beamforming problem is
transformed into a convex programming problem.
Then, a simplified version of the MPE beamforming is proposed which reduces
the exponential number of constraints in the MPE beamforming problem.
The simplified problem is also shown to be a convex programming problem.
The complexity of the simplified problem is further reduced by minimizing
a convex function which serves as an upper bound on the error probability.
Minimization of this upper bound results in
the introduction of a new metric, which is termed signal minus interference to
noise ratio (SMINR). It is shown that maximizing SMINR leads to a closed-form
expression for beamforming vectors
as well as improved performance over existing beamforming methods.
\end{abstract}

\section{Introduction}\label{secIntro_uplink}
\IEEEPARstart{I}{n} wireless communications, remarkable advantages
such as diversity, spatial multiplexing gain, and higher
throughput for single-user and multiuser systems are achieved
by using multiple transmit and receive antennas
\cite{Telatar99,Paulraj04,GesbertSPMag07,WeiYu04,Lim13}.
In a system which exploits antenna
arrays, space division multiple access (SDMA) techniques could be used to
obtain spatial multiplexing gain and to significantly
increase the achievable system throughput \cite{Veen88,LitvaBook96,
Sidiropoulos06}. 
Linear and nonlinear beamforming techniques
employed with an antenna array achieve spatial multiplexing by
separating users' signals transmitted simultaneously and
on the same carrier frequency, provided that
their channels are linearly independent
\cite{LitvaBook96,Gershman10,MajidICT11}. 
Classically, beamforming weights can be determined by
maximizing signal to noise ratio (SNR),
nulling the interference, i.e., zero-forcing (ZF) co-channel interference,
minimizing mean square error (MSE) between the desired signal and
the array output, maximizing signal to interference and noise ratio (SINR),
or minimizing the received signal variance while keeping the system
response distortionless (MVDR) \cite{GodaraPartI97,GodaraPartII97,Zhou12}.
However, in digital communications systems, the error probability
more closely reflects
actual quality of service (QoS)
\cite{Pados99,BlosteinCOM07,BlosteinVT07,Alouini13}.
Therefore, beamforming weights ought to be set with
the goal of directly minimizing the error probability.

Directly minimizing the error probability was considered in \cite{Barry97,Barry00}
for designing equalizers to combat intersymbol interference (ISI).
Later, this approach was adopted in a multiuser detection
(MUD) scenario to estimate the received signals by minimizing
the probability of error in a code division multiple access (CDMA)
system \cite{Pados99,Antoniou00,WeiAndChen07}.
Minimum probability of error (MPE) beamforming was
studied in \cite{Hanzo05,Hanzo08, MajidICT11} by extending the
ideas of MPE detection in CDMA systems and MPE equalization for ISI removal
to the problem of spatial multiplexing and receive beamforming.

It has been shown in \cite{MajidQBSC14,Hanzo05} that
MPE beamforming substantially outperforms ZF beamforming, minimum
mean square error (MMSE) beamforming, and other classical receive
beamforming methods. Nevertheless, the
probability of error function in a multiuser system is highly nonlinear and
suffers from the existence of numerous local minima \cite{Antoniou00,MajidQBSC14}.
This issue has been resolved for the special case
of MPE beamforming with binary phase shift keying (BPSK) by transforming
the nonconvex nonlinear MPE beamforming problem to a convex optimization problem
in \cite{MajidQBSC14}.
Nevertheless, the high computational complexity of the problem in \cite{MajidQBSC14}
is still an unresolved issue, besides its limitation to BPSK signalling.

In this paper, not only the idea of convex MPE beamforming is extended from BPSK to
general one-dimensional (1D) signalling, but also
the issue of high computational complexity is addressed.
First, we calculate the error probability of each pulse amplitude modulated user in
the uplink of a multiple access wireless system. Then, we formulate a beamforming
problem by minimizing the error probability of each user. The minimum
probability of error beamforming is then transformed to a convex optimization
problems with a unique solution. Next, the exponential complexity of the problem is
reduced by decreasing the number of constraints in the optimization.
Subsequently, we further reduce the complexity of the problem by minimizing an
upper bound on the error probability of each user. Finally, derived from the error probability,
a new metric is presented, which we term signal minus interference to noise ratio (SMINR).
Maximization of this metric results in a closed-form solution for the beamforming
weights of each user. It will be seen that maximizing SMINR also results in improved
performance compared to that of conventional ZF and MMSE beamforming.

The Internet of things (IoT) requires
simultaneous deployment of a massive number of low data rate devices \cite{Schwarz16,Nokia15}.
Therefore, a motivation for considering one-dimensional signalling is the
emergence of technologies such as IoT.
Moreover, power efficient BPSK, a special case of one-dimensional modulation, is a commonly
employed transmission mode in adaptive wireless systems such as IEEE 802.11a,n,ac,
when SNR is low \cite{Saeed14,Forney98}.

The rest of this paper is organized as follows: 
Section \ref{secSysModel_uplink} introduces the system model.
In Section \ref{secErrProb_uplink},
the exact error probability of each user in a vector multiple access channel is
calculated for one-dimensional modulation. 
In Section \ref{secConvexMpe_uplink}, the MPE
beamforming problem is transformed into a convex optimization problem.
Section \ref{secReducedMpe_uplink} reduces the complexity of convex MPE
beamforming problem and introduces the SMINR criterion and maximum SMINR beamforming.
Numerical results are presented in Section \ref{secResults_uplink}.
Finally, conclusions are drawn in Section \ref{secConclusion_uplink}.

The following mathematical notation is used throughout
the paper. Boldface upper case and lower case letters denote
matrices and vectors, respectively. The superscripts $(\cdot)^T$,
$(\cdot)^H$ denote the transpose and conjugate transpose, respectively.
The eigenvector corresponding to the maximum eigenvalue is denoted by
${\bf v}_\text{max}$. $\|\cdot\|_2$ denotes the $\ell_2$-norm.
$\Re\{\cdot\}$ and $\Im\{\cdot\}$ represent real and imaginary parts
of complex numbers/matrices, respectively.

\section{System Model}\label{secSysModel_uplink}
We consider a multiple access system supporting $K$ power-limited users, where each user
transmits a pulse amplitude modulated signal. It is assumed that users
are in the far-field region of a linear antenna array with $N$ elements.
It is further assumed that users transmit their
signals on the same carrier frequency, $f_c$. 
Baseband pulse amplitude modulated signal of user $k$ is represented by
\begin{equation}
    s_k(t;l_k) = A_k(l_k)g(t), \quad 1 \le k \le K,
\end{equation}
only over a real basis function,
where $A_k(l_k)$ takes on values from the set
\begin{equation}\label{pamSet_eq}
    \{(2l_k-1-L_k)d| ~1 \le l_k \le L_k \}
\end{equation}
with equal probability and $2d$ is the distance between adjacent signal constellation points.
Therefore, in vector space form, the transmitted signal of user $k$ is represented by
\begin{equation}
    s_k(l_k) = \sqrt{E_g} A_k(l_k), \quad 1 \le k \le K.
\end{equation}

Assuming fading channels and additive noise, the $N$-dimensional received
signal vector, ${\bf{r}}$ is represented by
\begin{equation}\label{rxSignal_eq}
    {\bf r} = {\bf H}{\bf s} + {\bf z},
\end{equation}
where ${\bf H} = [{\bf h}_1, \cdots, {\bf h}_K]$ and
${\bf h}_k$ is the $N$-dimensional channel vector between transmitter $k$
and the $N$ receive antennas.
The components of the ${\bf h}_k$s are assumed to follow an
independent identically distributed (i.i.d.) circularly
symmetric complex Gaussian (CSCG) distribution with zero mean and unit variance.
This channel model is valid for narrowband (frequency non-selective) systems if the
transmit and receive antennas are in
non line-of-sight rich-scattering environments with sufficient
antenna spacing \cite{PaulrajBook03,Mao12}.
In (\ref{rxSignal_eq}), ${\bf s} = [{s}_1, \ldots, {s}_K]^T$, where
$s_k = s_k(l_k)$ and the noise
${\bf z}$ is an $N$-dimensional vector, the elements of which are mutually
independent identically distributed CSCG random variables with
zero mean and variance $\sigma_z^2$.

Assuming linear processing at the receiver,
the array output for user $k$ can be written as a function of the received filter
of user $k$ as
\begin{align}\label{processedSignal_eq}
    \nonumber
    y_k &= {\bf w}_k {\bf r} =
    {\bf w}_k {\bf H} {\bf s} + {\bf w}_k {\bf z} =
    \sum_{j=1}^K{\bf w}_k {\bf h}_j s_j + {\bf w}_k {\bf z} \\
    \nonumber
    &= {\bf w}_k {\bf h}_k s_k + {\bf w}_k {\bf H}_{\bar k} {\bf s}_{\bar k} + {\bf w}_k {\bf z}\\
    &= {\bar y}_k + {z'_k} , \quad 1 \le k \le K,
\end{align}
where ${\bf w}_k$ is the $1 \times N$ complex-valued receive beamformer of user $k$, $z'_k$ is a
complex valued Gaussian noise with variance $\sigma_z^2 {\bf w}_k {\bf w}_k^H$,
${\bf s}_{\bar k} = [s_1, \cdots, s_{k-1}, s_{k+1}, \cdots, s_K]^T$, and
${\bf H}_{\bar k} = [{\bf h}_1, \cdots, {\bf h}_{k-1}, {\bf h}_{k+1}, \cdots, {\bf h}_K]$.

\section{Error Probability}\label{secErrProb_uplink}
Since the information signal is one-dimensionally modulated only on a real basis
function, without loss of generality,
the decision is performed only over the real part of the output of the
receive beamformer. 
For $1 \le k \le K$,
we consider the following decision rule for estimating the transmitted
symbols of user $k$:
\begin{align}\label{pamDecision_eq}
    \nonumber
    &{\hat s}_k \\
    &=\left\{ \begin{array}{lr}
    s_k(1) &  y_k^R \le \Re\{{\bf w}_k {\bf h}_k s_k(1) + {\bf w}_k {\bf h}_k d \sqrt{E_g}\}  \\[.1in]
    s_k(l_k)  & \begin{array}{r} \Re\{{\bf w}_k {\bf h}_k s_k(l_k) - {\bf w}_k {\bf h}_k d \sqrt{E_g}\} <
    y_k^R \\
    \le \Re\{{\bf w}_k {\bf h}_k s_k(l_k) + {\bf w}_k {\bf h}_k d \sqrt{E_g}\}; \\
    2 \le l_k \le L_k-1 \end{array} \\[.25in]
    s_k(L_k)  & y_k^R >  \Re\{{\bf w}_k {\bf h}_k s_k(L_k) - {\bf w}_k {\bf h}_k d \sqrt{E_g}\}
    \end{array} \right. ,
\end{align}
where the superscript $^R$ denotes the real part, i.e., $x^R={\Re}\{x\}$.

The error probability of user $k$ is expressed as
\begin{align}\label{uplinkPeUser_eq}
    \nonumber
    &P_{e_k} \!=\! \sum_{l_k=1}^{L_k} \! P(l_k) P_{e_k}(l_k) \!=\!
    \frac{1}{L_k} \! \sum_{l_k=1}^{L_k} \! P({\hat s}_k \!\neq\! s_k(l_k)| s_k \!=\! s_k(l_k)) \\
    \nonumber
    &= \! \frac{1}{L_k} \!
    \left[\!P(y_k^R \!>\! \Re\{{\bf w}_k {\bf h}_k s_k(1) \!+\!
    {\bf w}_k {\bf h}_k d \sqrt{E_g}\} | s_k \!=\! s_k(1)) \right. \\
    \nonumber
    &\!+\!\! \sum_{l_k\!=\!2}^{L_k\!-\!1} \!\!
    P(|y_k^R \!\!-\! \Re\{{\bf w}_k {\bf h}_k s_k(l_k\!)\}| \!\!>\!\!
    \Re\{{\bf w}_k {\bf h}_k d \sqrt{\!E_g}\} \!| s_k \!\!=\!\! s_k(l_k)) \\
    &+\! \left. P(y_k^R \!\le\! \Re\{{\bf w}_k {\bf h}_k s_k(L_k) \!-\!
    {\bf w}_k {\bf h}_k d \sqrt{E_g}\}| s_k \!=\! s_k(L_k)) \!\right]
\end{align}
where $P(\cdot)$ is the probability of an event and
$P(l_k) = P(s_k = s_k(l_k)) = \frac{1}{L_k}$, i.e., the transmitted PAM signal $s_k$
takes its values from the set (\ref{pamSet_eq}) with equal probability.
The error probability of user $k$, given $s_k(l_k)$ is transmitted,
is denoted by $P_{e_k}(l_k)$.
It should be remarked that assuming uniform (as we did) rather than Gaussian
distribution over signal sets, although more practical,
causes an asymptotic loss in throughput
which could be compensated to some extent by using constellation shaping
techniques \cite{Forney98}.

To calculate the error probability (\ref{uplinkPeUser_eq}), first we need to find
the probability density function (pdf) of $y_k^R$ conditioned on $s_k$, namely,
$p(y_k^R|s_k)$. Let us denote the number of possible symbol sequences of
all $K$ users in one transmission by $N_{b}=\prod_{k=1}^K L_k$, i.e.,
there could be $N_b$ different possible sets of $K$-tuple symbols
${\bf s}$ for $K$ users. Moreover, let
$N_{p_k} = \prod_{\substack{j=1\\ j \neq k}}^K L_j$ denote the number of possible vector
of symbols for transmission if the transmitted symbol of user $k$ is already known, i.e.,
there could be $N_{p_k}$ different possible sets of $K-1$-tuple
symbols ${\bf s}_{\bar k}(b),~ 1 \leq b \leq N_{p_k},$ for $K-1$ users.
Using equal probability for transmission of PAM constellation points, and
Gaussian output noise $\Re\{z'_k\}$, we have
\begin{align}\label{pdfUplink_eq}
    \nonumber
    &p(y_k^R | s_k = s_k(l_k))
    = \frac{1}{N_{p_k}} \sum_{\forall {\bf s}_{\bar k}}
    p(y_k^R | s_k = s_k(l_k), {\bf s}_{\bar k}) \\
    \nonumber
    &=\frac{1}{N_{p_k}} \sum_{b = 1}^{N_{p_k}}
    p(y_k^R | s_k = s_k(l_k), {\bf s}_{\bar k} = {\bf s}_{\bar k}(b)) \\
    &= \frac{1}{N_{p_k}}
    \sum_{b = 1}^{N_{p_k}} \frac{1}{\sqrt{\pi \sigma_z^2 {\bf w}_k {\bf w}_k^H}}
    \exp{-\frac{(y_k^R - {\bar y}_k^R(l_k,b))^2} {\sigma_z^2 \|{\bf w}_k\|_2^2}},
\end{align}
where in the first equality the total probability theorem is used to
condition the conditional output probability of user $k$ over all $N_{p_k}$
possible symbol assignment of the transmitted symbols ${\bf s}_{\bar k}$.
Also ${\bar y}_{k}^R(l_k,b) = {\Re}\{{\bar y}_{k}\}$ when
$s_k = s_k(l_k)$ and ${\bf s}_{\bar k} = {\bf s}_{\bar k} (b)$, i.e.,
\begin{align}
    \nonumber
    {\bar y}_{k}^R(l_k, b) = \Re\{{\bf w}_k {\bf h}_k s_k(l_k)  +
    {\bf w}_k {\bf H}_{\bar k} {\bf s}_{\bar k}(b)\},\\
    \quad 1 \leq l_k \leq L_k, ~1 \leq b \leq N_{p_k}.
\end{align}

Having the conditional pdf of $y_k^R$ as in (\ref{pdfUplink_eq}), each of the three
terms in the last equality of (\ref{uplinkPeUser_eq}) can be calculated as follows:
\begin{align}\label{uplinkPeUserPart1_eq}
    \nonumber
    & P\left(y_k^R > \Re\{{\bf w}_k {\bf h}_k s_k(1) +
    {\bf w}_k {\bf h}_k d \sqrt{E_g}\} | s_k = s_k(1)\right) \\
    \nonumber
    &= \int_{\Re\{{\bf w}_k {\bf h}_k s_k(1) +
    {\bf w}_k {\bf h}_k d \sqrt{E_g}\}}^{\infty}
    \frac{1}{N_{p_k}}
    \sum_{b = 1}^{N_{p_k}} \frac{1}{\sqrt{\pi \sigma_z^2 {\bf w}_k {\bf w}_k^H}} \\
    \nonumber
    &
    \times \exp{-\frac{(y_k^R - {\bar y}_k^R(1,b))^2} {\sigma_z^2 \|{\bf w}_k\|_2^2}}
    dy_k^R \\
    \nonumber
    &= \frac{1}{N_{p_k}}
    \sum_{b = 1}^{N_{p_k}} Q\left(\frac{ \Re\{{\bf w}_k {\bf h}_k d \sqrt{E_g} -
    {\bf w}_k {\bf H}_{\bar k} {\bf s}_{\bar k}(b)\}} {\frac{\sigma_z}{\sqrt{2}} \|{\bf w}_k\|_2}
    \right), \\
    &1 \le k \le K,
\end{align}
where the $Q$-function is defined as
$Q(x)=\frac{1}{\sqrt{2\pi}}\int_x^{\infty} {e^{-\frac{u^2}{2}}du}$, and we used the
property
\begin{equation}\label{qFuncProperty1_eq}
    \frac{1}{\sqrt{2\pi\sigma^2}} \int_x^\infty \exp\left(-\frac{(u-\bar u)^2}{2\sigma^2}
    \right) \, du = Q(\frac{x-\bar u}{\sigma}).
\end{equation}
The third part of the last equality in (\ref{uplinkPeUser_eq}) is calculated as
\begin{align}\label{uplinkPeUserPart3_eq}
    \nonumber
    &P\left(y_k^R \le \Re\{{\bf w}_k {\bf h}_k s_k(L_k) -
    {\bf w}_k {\bf h}_k d \sqrt{E_g}\}| s_k = s_k(L_k)\right) \\
    \nonumber
    &= \int_{-\infty}^{\Re\{{\bf w}_k {\bf h}_k s_k(L_k) -
    {\bf w}_k {\bf h}_k d \sqrt{E_g}\}}
    \frac{1}{N_{p_k}}
    \sum_{b = 1}^{N_{p_k}} \\
    \nonumber
    &\frac{1}{\sqrt{\pi \sigma_z^2 {\bf w}_k {\bf w}_k^H}}
    \exp{-\frac{(y_k^R - {\bar y}_k^R(L_k,b))^2} {\sigma_z^2 \|{\bf w}_k\|_2^2}}
    dy_k^R \\
    \nonumber
    &= \frac{1}{N_{p_k}}
    \sum_{b = 1}^{N_{p_k}} Q\left(\frac{ \Re\{{\bf w}_k {\bf h}_k d \sqrt{E_g} +
    {\bf w}_k {\bf H}_{\bar k} {\bf s}_{\bar k}(b)\}} {\frac{\sigma_z}{\sqrt{2}} \|{\bf w}_k\|_2}
    \right), \\
    &1 \le k \le K,
\end{align}
where we used the property
\begin{equation}\label{qFuncProperty2_eq}
    \frac{1}{\sqrt{2\pi\sigma^2}} \int_{-\infty}^x \exp\left(-\frac{(u-\bar u)^2}{2\sigma^2}
    \right) \, du = Q(\frac{\bar u - x}{\sigma}).
\end{equation}
The second part of the last equality in (\ref{uplinkPeUser_eq}) is calculated as
\begin{align}\label{uplinkPeUserPart2_eq}\nonumber
    & P\left(|y_k^R - \Re\{{\bf w}_k {\bf h}_k s_k(l_k)\}| \!>\!
    \Re\{{\bf w}_k {\bf h}_k d \sqrt{E_g}\} | s_k \!=\! s_k(l_k)\right) \\
    \nonumber
    &\!=\! P\left(y_k^R \!-\! \Re\{{\bf w}_k {\bf h}_k s_k(l_k)\} \!>\!
    \Re\{{\bf w}_k {\bf h}_k d \sqrt{E_g}\} | s_k \!=\! s_k(l_k)\right) \\
    \nonumber
    &\!+\! P\!\left(\!y_k^R \!-\! \Re\{\!{\bf w}_k {\bf h}_k s_k(l_k)\} \!<\!\! -
    \Re\{{\bf w}_k {\bf h}_k d \sqrt{E_g}\}| s_k \!=\! s_k(l_k)\right) \\
    \nonumber
    &\!=\! \frac{1}{N_{p_k}} \sum_{b = 1}^{N_{p_k}}
    \left[Q(\frac{ \Re\{{\bf w}_k {\bf h}_k d \sqrt{E_g} -
    {\bf w}_k {\bf H}_{\bar k} {\bf s}_{\bar k}(b)\}} {\frac{\sigma_z}{\sqrt{2}} \|{\bf w}_k\|_2}) \right. \\
    \nonumber
    &\!+
    \left. Q(\frac{ \Re\{{\bf w}_k {\bf h}_k d \sqrt{E_g} \!+\!
    {\bf w}_k {\bf H}_{\bar k} {\bf s}_{\bar k}(b)\}} {\frac{\sigma_z}{\sqrt{2}} \|{\bf w}_k\|_2}
    ) \right]\!, \\
    & 2 \le l_k \le L_k-2, \quad 1 \le k \le K,
\end{align}
where (\ref{qFuncProperty1_eq}) and (\ref{qFuncProperty2_eq}) are used.
Finally, using (\ref{uplinkPeUser_eq}), (\ref{uplinkPeUserPart1_eq}),
(\ref{uplinkPeUserPart3_eq}), and (\ref{uplinkPeUserPart2_eq}) yields
the error probability of user $k$:
\begin{align}\label{userError2QFunc_eq}
    \nonumber
    &P_{e_k} = \frac{L_k-1}{N_b} \sum_{b=1}^{N_{p_k}} \\
    \nonumber
    &\left[Q\left(\frac{ \Re\{{\bf w}_k {\bf h}_k d \sqrt{E_g} -
    {\bf w}_k {\bf H}_{\bar k} {\bf s}_{\bar k}(b)\}} {\frac{\sigma_z}{\sqrt{2}} \|{\bf w}_k\|_2}\right) \right. \\
    &+ \left.Q\left(\frac{ \Re\{{\bf w}_k {\bf h}_k d \sqrt{E_g} +
    {\bf w}_k {\bf H}_{\bar k} {\bf s}_{\bar k}(b)\}} {\frac{\sigma_z}{\sqrt{2}} \|{\bf w}_k\|_2}
    \right) \right].
\end{align}

To proceed further, the following property is required:
\begin{property}\label{propertyEqualQ}
    For a given user
    \begin{align}
        \nonumber
        &\sum_{b=1}^{N_{p_k}} Q\left(\frac{ \Re\{{\bf w}_k {\bf h}_k d \sqrt{E_g} -
        {\bf w}_k {\bf H}_{\bar k} {\bf s}_{\bar k}(b)\}}
        {\frac{\sigma_z}{\sqrt{2}} \|{\bf w}_k\|_2}\right) \\
        &=\sum_{b=1}^{N_{p_k}} Q\left(\frac{ \Re\{{\bf w}_k {\bf h}_k d \sqrt{E_g} +
        {\bf w}_k {\bf H}_{\bar k} {\bf s}_{\bar k}(b)\}}
        {\frac{\sigma_z}{\sqrt{2}} \|{\bf w}_k\|_2}\right).
    \end{align}
\end{property}
\begin{IEEEproof}
    The pulse amplitude modulated signal constellation of each user is symmetric
    about zero. 
    Therefore, for every given $b=b_1$ there exists a $b=\bar b_1$ such that
    ${\bf s}_{\bar k}(b_1) = -{\bf s}_{\bar k}(\bar b_1)$.
    Therefore,
    \begin{align}
        \nonumber
        &Q(\frac{ \Re\{{\bf w}_k {\bf h}_k d \sqrt{E_g} -
        {\bf w}_k {\bf H}_{\bar k} {\bf s}_{\bar k}(b_1)\}}
        {\frac{\sigma_z}{\sqrt{2}} \|{\bf w}_k\|_2}) \\
        &=
        Q(\frac{ \Re\{{\bf w}_k {\bf h}_k d \sqrt{E_g} +
        {\bf w}_k {\bf H}_{\bar k} {\bf s}_{\bar k}(\bar b_1)\}}
        {\frac{\sigma_z}{\sqrt{2}} \|{\bf w}_k\|_2}).
    \end{align}
    Hence,
    \begin{align}
        \nonumber
        &\sum_{b=1}^{N_{p_k}} Q\left(\frac{ \Re\{{\bf w}_k {\bf h}_k d \sqrt{E_g} -
        {\bf w}_k {\bf H}_{\bar k} {\bf s}_{\bar k}(b)\}}
        {\frac{\sigma_z}{\sqrt{2}} \|{\bf w}_k\|_2}\right) \\
        \nonumber
        &=\sum_{\bar b=1}^{N_{p_k}} Q\left(\frac{ \Re\{{\bf w}_k {\bf h}_k d \sqrt{E_g} +
        {\bf w}_k {\bf H}_{\bar k} {\bf s}_{\bar k}(\bar b)\}}
        {\frac{\sigma_z}{\sqrt{2}} \|{\bf w}_k\|_2}\right) \\
        &=\sum_{b=1}^{N_{p_k}} Q\left(\frac{ \Re\{{\bf w}_k {\bf h}_k d \sqrt{E_g} +
        {\bf w}_k {\bf H}_{\bar k} {\bf s}_{\bar k}(b)\}}
        {\frac{\sigma_z}{\sqrt{2}} \|{\bf w}_k\|_2}\right).
    \end{align}
\end{IEEEproof}
\noindent Using Property \ref{propertyEqualQ}, the error probability of user $k$
(\ref{userError2QFunc_eq}) can be expressed as
\begin{align}\label{userError_eq}
    \nonumber
    P_{e_k} = &\frac{2(L_k-1)}{L_k N_p} \sum_{b=1}^{N_{p_k}}
    Q\left(\frac{ \Re\{{\bf w}_k {\bf h}_k d \sqrt{E_g} +
    {\bf w}_k {\bf H}_{\bar k} {\bf s}_{\bar k}(b)\}}
    {\frac{\sigma_z}{\sqrt{2}} \|{\bf w}_k\|_2}\right) \\
    =&\frac{2(L_k-1)}{L_k N_p} \sum_{b=1}^{N_{p_k}}
    Q\left(\frac{ \Re\{{\bf w}_k {\bf h}_k d \sqrt{E_g} -
    {\bf w}_k {\bf H}_{\bar k} {\bf s}_{\bar k}(b)\}}
    {\frac{\sigma_z}{\sqrt{2}} \|{\bf w}_k\|_2}\right).
\end{align}

\subsection{Minimum Probability of Error (MPE) Receive Beamforming}\label{subSecMpe_uplink}
Knowing each user's modulation type, its receive
beamforming weights can be calculated by minimizing
its probability of error. Therefore, the minimum probability of error (MPE) beamforming
weights are the solutions to the following optimization problem:
\begin{equation}\label{mpeOriginal_eq}
    {\bf {w}}_{k_{\text{MPE}}} = \argmin_{{\bf w}_k} P_{e_k} , \quad 1 \le k \le K,
\end{equation}
where $P_{e_k}$ is the error probability of user $k$ defined in (\ref{userError_eq}).
As it can be seen from (\ref{mpeOriginal_eq}) and (\ref{userError_eq}), the objective
function of MPE beamforming problem, i.e., the error probability of user $k$ in the
uplink of a multiuser system is a non-convex and nonlinear function of beamforming
vector ${\bf w}_k$.
While in general the non-convex and nonlinear optimization problem
(\ref{mpeOriginal_eq}) can be solved by using exhaustive (brute force) search to achieve a global
minimum, its computational complexity is prohibitive \cite{NocedalBook06}.
On the other hand, gradient-based optimization algorithms such as BFGS \cite{NocedalBook06} can
at best guarantee local stationary points.
In our simulations, it was observed that if the initial point for the
gradient-based optimization algorithm is not chosen appropriately,
the algorithm converges to a drastically poor solution.
Therefore, a more practical approach is necessary to
solve (\ref{mpeOriginal_eq}).

\section{Convex Optimization Based MPE Receive Beamforming}\label{secConvexMpe_uplink}
In this section, similar to \cite{Antoniou00} the MPE receive beamforming problem is
transformed into a convex optimization problem with a unique solution which
can be obtained by conventional convex programming algorithms such as interior point
methods \cite{BoydBook04}.

First, it should be remarked that ideally the error probability of all users in a
MAC channel should approach zero when the transmit power of users approach infinity.
To this end, we define the error floor as follows:
\begin{definition}
    If the transmit powers of all users approach infinity and yet the average
    error probability cannot approach zero, the value of the tight lower
    bound on the average of the error probability is called the error floor.
\end{definition}
\begin{proposition}\label{propositionPositiveArgQ}
    For the users not to have an error floor, it is necessary
    for ${\bf w}_k,~1 \le k \le K$ to comply with the following constraints:
    \begin{equation}\label{PositiveArgQ_eq}
        { \Re\{{\bf w}_k {\bf h}_k d \sqrt{E_g} -
        {\bf w}_k {\bf H}_{\bar k} {\bf s}_{\bar k}(b)\}} \ge 0, \quad 1 \leq b \leq N_{p_k}.
    \end{equation}
\end{proposition}
\begin{IEEEproof}
    See Appendix \ref{subSecAppendPropositionPositiveArgQ}.
\end{IEEEproof}

\begin{property}\label{propertyScaleQ}
The error probability in (\ref{userError_eq}) is invariant to the scaling of ${\bf w}_k$
by a positive constant.
\end{property}
\begin{IEEEproof}
    See Appendix \ref{subSecAppendPropertyScaleQ}.
\end{IEEEproof}

From Proposition \ref{propositionPositiveArgQ} and Property \ref{propertyScaleQ}, it becomes
clear that when no error floor exists, without loss of
generality, the constraints $\|{\bf w}_k\|=1$ and (\ref{PositiveArgQ_eq}) can
be added to the optimization problem (\ref{mpeOriginal_eq}). Therefore, the
MPE beamforming problem could be rewritten as follows.
\begin{subequations}\label{mpeConstrained_eq}
\begin{align}
    \label{mpeConstrained_subeq1}
    &{\bf {w}}_{k_{\text{MPE}}} = \argmin_{{\bf w}_k} P_{e_k} \\
    \label{mpeConstrained_subeq2}
    &\text{subject to} \quad \|{\bf w}_k\|_2 = 1 \\
    \label{mpeConstrained_subeq3}
    &\quad\quad\quad\quad\quad  \Re\{{\bf w}_k {\bf h}_k d \sqrt{E_g} -
    {\bf w}_k {\bf H}_{\bar k} {\bf s}_{\bar k}(b)\} \!\ge\! 0, 1 \!\leq\! b \!\leq\! N_{p_k}.
\end{align}
\end{subequations}

To solve the optimization problem (\ref{mpeConstrained_eq}), we have
\begin{theorem}\label{theoremUniqueGlobalMinimizer}
    If the constraints (\ref{mpeConstrained_subeq2}) and
    (\ref{mpeConstrained_subeq3}) in the minimization problem (\ref{mpeConstrained_eq}) are
    satisfied, any local minimizer of error probability function $P_{e_k}$ (\ref{userError_eq}),
    i.e., the objective function of the optimization problem (\ref{mpeConstrained_eq}), is
    also a global minimizer. Moreover, the global minimizer is unique.
\end{theorem}
\begin{IEEEproof}
    See Appendix \ref{subSecAppendTheoremUniqueGlobalMinimizer}.
\end{IEEEproof}

Although Theorem \ref{theoremUniqueGlobalMinimizer} shows that the constrained MPE problem
(\ref{mpeConstrained_eq}) has a unique global minimizer, (\ref{mpeConstrained_eq})
is not in the form of a standard convex
programming problem. However, the constrained optimization problem (\ref{mpeConstrained_eq}) can be
rewritten as follows:
\begin{subequations}\label{mpeConstrained2_eq}
\begin{align}
    \label{mpeConstrained2_subeq1}
    & \min_{{\bf w}_k}
    \frac{2(L_k-1)}{N_b} \sum_{b=1}^{N_{p_k}}
    Q\left(\frac{ \Re\{{\bf w}_k {\bf h}_k d \sqrt{E_g} -
    {\bf w}_k {\bf H}_{\bar k} {\bf s}_{\bar k}(b)\}}
    {\frac{\sigma_z}{\sqrt{2}} }\right) \\
    \label{mpeConstrained2_subeq2}
    &\text{subject to} \quad \|{\bf w}_k\|_2 = 1, \\
    \label{mpeConstrained2_subeq3}
    &\quad\quad\quad\quad\quad  \Re\{{\bf w}_k {\bf h}_k d \sqrt{E_g} -
    {\bf w}_k {\bf H}_{\bar k} {\bf s}_{\bar k}(b)\} \!\ge\! 0, 1 \!\leq\! b \!\leq\! N_{p_k},
\end{align}
\end{subequations}
which is the result of considering the equality constraint (\ref{mpeConstrained_subeq2})
in the denominator of the $Q$-functions in (\ref{userError_eq}).
However, the constrained problem (\ref{mpeConstrained2_eq}) is not a
convex problem either, since the feasible region
is not a convex set,
which is due to the fact that (\ref{mpeConstrained2_subeq2}) is not a convex set.
However, by transforming (\ref{mpeConstrained2_subeq2}) to
\begin{equation}\label{insideSphere_eq}
\|{\bf w}_k\|_2 \leq 1 ,
\end{equation}
the feasible region and therefore the optimization problem will
become convex\footnote{The set (\ref{insideSphere_eq}) is a convex set, since it represents
the interior and boundary of an $N$-dimensional sphere.}.
Furthermore, 
the
constraint set defined by (\ref{insideSphere_eq}) is an active set \cite{NocedalBook06}.
In other words, the
minimizer always satisfies the constraint $\|{\bf w}_k\|_2=1$,
because for $\|{\bf w}_k\|_2 < 1$, there always
exists a ${\bf \hat w}_k = \frac{{\bf w}_k} {\|{\bf w}_k\|_2}$ for
which $P_{e_k}({\bf \hat w}_k) < P_{e_k}({\bf w}_k)$.
Therefore, the MPE receive beamforming problem can be cast into the following
convex optimization problem with a unique global minimizer:
\begin{subequations}\label{mpeConstrained3_eq}
\begin{align}
    \label{mpeConstrained3_subeq1}
    & \min_{{\bf w}_k}
    \frac{2(L_k-1)}{N_b} \sum_{b=1}^{N_{p_k}}
    Q\left(\frac{ \Re\{{\bf w}_k {\bf h}_k d \sqrt{E_g} -
    {\bf w}_k {\bf H}_{\bar k} {\bf s}_{\bar k}(b)\}}
    {\frac{\sigma_z}{\sqrt{2}} }\right) \\
    \label{mpeConstrained3_subeq2}
    &\text{subject to} \quad \|{\bf w}_k\|_2 \le 1 \\
    \label{mpeConstrained3_subeq3}
    &\quad\quad\quad\quad\quad  \Re\{{\bf w}_k {\bf h}_k d \sqrt{E_g} -
    {\bf w}_k {\bf H}_{\bar k} {\bf s}_{\bar k}(b)\} \!\ge\! 0, 1 \!\leq\! b \!\leq\! N_{p_k}.
\end{align}
\end{subequations}
Problem (\ref{mpeConstrained3_eq}) can then be solved by conventional convex
programming methods. For example by using the interior point
methods, the complexity would be of polynomial
order with respect to $N$ \cite{NocedalBook06}.

It is worth noting
that the convex constrained problem (\ref{mpeConstrained3_eq})
only has a solution when the
set of constraints is feasible. If users in the original MPE
beamforming problem (\ref{mpeOriginal_eq}) suffer from the existence
of the error floor, it means that at least one of the constraints defined in
(\ref{mpeConstrained3_subeq3}) do not hold for one of the users.
Therefore, the set of constraints
defined by (\ref{mpeConstrained3_subeq2}) and (\ref{mpeConstrained3_subeq3})
is empty for this user.
In other words, the constrained
problem does not have any solution for this user. It should be mentioned that if
such an instance occurs, i.e., when users suffer from the existence
of the error floor, it is inherently impossible for at least one of the users to be met by
an acceptable quality of service using linear beamforming methods.
This means that the received signals of such a user are not linearly separable
using linear beamforming methods.

\section{Reduced-Complexity Convex MPE Beamforming}\label{secReducedMpe_uplink}
Although (\ref{mpeConstrained3_eq}) is a convex optimization problem with low
complexity in the number of receive antennas $N$, the number of constraints in
(\ref{mpeConstrained3_subeq3}) and the number of summations in (\ref{mpeConstrained3_subeq1}) are
of the order of $N_{p_k} = \prod_{\substack{j = 1\\j \ne k}}^K L_j$, i.e.,
exponential in the number of users $K$ and polynomial in the modulation order $L$.

\begin{proposition}\label{propositionSufficient}
    A necessary and sufficient condition for all $N_{p_k}$ constraints of (\ref{mpeConstrained3_subeq3})
    to hold is
    \begin{equation}\label{propositionSufficient_eq}
        \Re\{{\bf w}_k {\bf h}_k d \sqrt{E_g} \} - \sum_{\substack{j=1\\j\ne k}}^K
        |\Re\{{\bf w}_k {\bf h}_j s_j(L_j)\}| \ge 0.
    \end{equation}
\end{proposition}
\begin{IEEEproof}
    See Appendix \ref{subSecAppendPropositionSufficient}.
\end{IEEEproof}

Replacing all $N_{p_k}$ constraints of (\ref{mpeConstrained3_subeq3})
with (\ref{propositionSufficient_eq}), the MPE receive beamforming problem
(\ref{mpeConstrained3_eq}) is equivalently converted to
\begin{subequations}\label{mpeReduced_eq}
\begin{align}
    \label{mpeReduced_subeq1}
    & \min_{{\bf w}_k}
    \frac{2(L_k-1)}{N_b} \sum_{b=1}^{N_{p_k}}
    Q\left(\frac{ \Re\{{\bf w}_k {\bf h}_k d \sqrt{E_g} -
    {\bf w}_k {\bf H}_{\bar k} {\bf s}_{\bar k}(b)\}}
    {\frac{\sigma_z}{\sqrt{2}} }\right) \\
    \label{mpeReduced_subeq2}
    &\text{subject to} \quad \|{\bf w}_k\|_2 \le 1 \\
    \label{mpeReduced_subeq3}
    &\quad\quad\quad\quad\quad~  \Re\{{\bf w}_k {\bf h}_k d \sqrt{E_g} \} -
    \sum_{\substack{j=1\\j\ne k}}^K
    |\Re\{{\bf w}_k {\bf h}_j s_j(L_j)\}| \ge 0,
\end{align}
\end{subequations}
with reduced complexity.

\begin{claim}\label{claimConvexSet}
    Constraint set (\ref{mpeReduced_subeq3}) is a convex set.
\end{claim}
\begin{IEEEproof}
    See Appendix \ref{subSecAppendClaimConvexSet}.
\end{IEEEproof}
\noindent Based on Claim \ref{claimConvexSet} and the discussion in Section \ref{secConvexMpe_uplink},
it can be inferred that the reduced MPE beamforming problem (\ref{mpeReduced_eq}) is a convex
problem which can be solved using conventional convex programming methods
\cite{BoydBook04}.


\subsection{Maximum Signal Minus Interference to Noise Ratio (SMINR)
Beamforming - Amplitude Version}\label{subSecSminrAmp_uplink}
To further reduce the complexity of (\ref{mpeReduced_eq}), we replace
the summation of $N_{p_k}$ terms in objective function (\ref{mpeReduced_subeq1}) by
a single term which serves as an upper bound on the objective function.
\begin{claim}\label{claimErrorUpperBound}
    The following expression is an upper bound on the error probability of user $k$
    and therefore on the objective function (\ref{mpeReduced_subeq1}):
    \begin{align}\label{userErrorUpperBound_eq}
        \nonumber
        &\tilde P_{e_k}^{\text{Up}} =
        \frac{2(L_k-1)}{L_k} \\
        &\times\! Q\left(\frac{ \Re\{{\bf w}_k {\bf h}_k d \sqrt{E_g} \} -
        \sum_{\substack{j=1\\j\ne k}}^K
        |\Re\{{\bf w}_k {\bf h}_j s_j(L_j)\}|}
        {\frac{\sigma_z}{\sqrt{2}} }\right)\!.
    \end{align}
\end{claim}
\begin{IEEEproof}
    Considering that the $Q$-function is decreasing for nonnegative
    arguments, and using (\ref{qFuncArgLowerBound_eq}) in
    Appendix \ref{subSecAppendPropositionSufficient}, it can be easily
    shown that (\ref{userErrorUpperBound_eq}) is an upper bound on the
    error probability of user $k$ and also an upper bound on
    the objective function (\ref{mpeReduced_subeq1}).
\end{IEEEproof}

Using (\ref{userErrorUpperBound_eq}), the beamforming problem is formulated
by minimizing the upper bound
on the error probability of each user:
\begin{subequations}\label{mpeReduced2_eq}
\begin{align}
    \label{mpeReduced2_subeq1}
    & {\bf {w}}_{k_\text{SMINR\_Amp}} = \argmin_{{\bf w}_k} \tilde P_{e_k}^{\text{Up}} \\
    \label{mpeReduced2_subeq2}
    &\text{subject to} \quad \|{\bf w}_k\|_2 \le 1 \\
    \label{mpeReduced2_subeq3}
    &\quad\quad\quad\quad\quad~  \Re\{{\bf w}_k {\bf h}_k d \sqrt{E_g} \} -
    \sum_{\substack{j=1\\j\ne k}}^K
    |\Re\{{\bf w}_k {\bf h}_j s_j(L_j)\}| \ge 0.
\end{align}
\end{subequations}
Since the argument of the $Q$-function in (\ref{userErrorUpperBound_eq}) is
constrained to be nonnegative by (\ref{mpeReduced2_subeq3}),
and the $Q$-function is a decreasing function for
nonnegative arguments, problem (\ref{mpeReduced2_eq}) is equivalent to
\begin{subequations}\label{sminrBf_eq}
\begin{align}
    \label{sminrBf_subeq1}
    & {\bf w}_{k_\text{SMINR\_Amp}} = \argmax_{{\bf w}_k}
    \text{ SMINR}_\text{Amp} \\
    \label{sminrBf_subeq2}
    &\text{subject to} \quad \|{\bf w}_k\|_2 \le 1 \\
    \label{sminrBf_subeq3}
    &\quad\quad\quad\quad\quad~  \Re\{{\bf w}_k {\bf h}_k d \sqrt{E_g} \} -
    \sum_{\substack{j=1\\j\ne k}}^K
    |\Re\{{\bf w}_k {\bf h}_j s_j(L_j)\}| \ge 0,
\end{align}
\end{subequations}
where
\begin{equation}\label{sminr_eq}
    \text{SMINR}_\text{Amp} = \frac{ \Re\{{\bf w}_k {\bf h}_k d \sqrt{E_g} \} -
    \sum_{\substack{j=1\\j\ne k}}^K |\Re\{{\bf w}_k {\bf h}_j s_j(L_j)\}|}
    {\frac{\sigma_z}{\sqrt{2}} }.
\end{equation}
The objective function of (\ref{sminrBf_eq}), i.e., (\ref{sminr_eq})
can be interpreted as the ratio of half the distance between
two received signal constellation points minus the maximum amplitude of the
interference from all other users relative to the noise standard deviation (square root of noise variance).
We name this objective function signal minus interference
to noise ratio (SMINR), amplitude-based version. Maximum SMINR beamforming-amplitude version
(SMINR-Amp)
(\ref{sminrBf_eq}) is a low-complexity convex optimization problem since the constraints
(\ref{sminrBf_subeq3}) and (\ref{sminrBf_subeq2}) are convex sets as
discussed in Claim \ref{claimConvexSet} and the discussion in
Section \ref{secConvexMpe_uplink}, respectively; also it can easily be
shown that the objective function of (\ref{sminrBf_eq}) is a concave
function by using the definition of convex and concave functions \cite{BoydBook04}.

\subsection{Heuristic Maximum SMINR Beamforming}\label{subSecSminrPower_uplink}
Although (\ref{sminrBf_eq}) is a low-complexity convex optimization problem, it still
needs to be solved using numerical optimization methods.
We next aim to formulate a similar problem to (\ref{sminrBf_eq}) that can be dealt with
analytically to obtain a closed-form solution.
To this end, we define a power-based version of SMINR as follows:
\begin{equation}\label{sminrPower_eq}
    \text{SMINR} \triangleq \frac{ (\Re\{{\bf w}_k {\bf h}_k d \sqrt{E_g}\})^2 -
    \sum_{\substack{j=1\\j\ne k}}^K (\Re\{{\bf w}_k {\bf h}_j s_j(L_j)\})^2}
    {\frac{\sigma_z^2}{2} }.
\end{equation}
The following optimization problem is then introduced:
\begin{subequations}\label{sminrBf2_eq}
\begin{align}
    \label{sminrBf2_subeq1}
    & {\bf w}_{k_\text{SMINR}} = \argmax_{{\bf w}_k}
    \text{ SMINR} \\
    \label{sminrBf2_subeq2}
    &\text{subject to} \quad \|{\bf w}_k\|_2 = 1.
\end{align}
\end{subequations}
Power-based SMINR (\ref{sminrPower_eq}), which for simplicity is termed
SMINR henceforward, is differentiable with respect to ${\bf w}_k$
and method of Lagrange multipliers can be adopted to solve the corresponding optimization
problem.
Writing the Lagrangian and using Wirtinger calculus \cite{Brandwood83,GesbertTSP07,Koivunen10}
to set the gradient of the
Lagrangian with respect to ${\bf w}_k$ to zero, the following equation is obtained:
\begin{equation}\label{entangledComplexSminr_eq}
    d^2E_g\Re\{{\bf w}_k {\bf h}_k\} {\bf h}_k^T \!-\! \sum_{\substack{j=1\\j\ne k}}^K
    s_j^2(L_j) \Re\{{\bf w}_k {\bf h}_j\} {\bf h}_j^T \!+\!
    \mu \frac{\sigma_z^2}{2} {\bf w}_k^* \!=\! {\bf 0},
\end{equation}
where $\mu$ is the Lagrange multiplier. Unfortunately, as can be seen from
(\ref{entangledComplexSminr_eq}), ${\bf w}_k$ and ${\bf w}_k^*$ are
coupled in a way that a closed-form solution cannot be obtained. Since the objective
here was to find a closed-form solution for beamforming
vectors, pursuing this approach is not of interest.

To tackle the coupling issue, the following transformations are employed
for $1 \le k \le K$:
\begin{align}\label{transformations_eq}
    \nonumber
    &{\bf w}_k \in \mathbb{C}^{1\times N} \xrightarrow{\mathcal{T}_1}
    {\bf\bar w}_k = \left[\begin{array}{cc}
    \Re\{{\bf w}_k\} & \Im\{{\bf w}_k\} \end{array} \right] \in
    \mathbb{R}^{1\times 2N}, \\
    &{\bf h}_k \in \mathbb{C}^{N \times 1} \xrightarrow{\mathcal{T}_2} {\bf\tilde h}_k =
    \left[\begin{array}{c}
    \Re\{{\bf h}_k\} \\ -\Im\{{\bf h}_k\} \end{array} \right]
    \in \mathbb{R}^{2N\times 1} .
\end{align}
Using (\ref{transformations_eq}), maximum SMINR receive beamforming
(\ref{sminrBf2_eq}) is reformulated as
\begin{subequations}\label{sminrBf3_eq}
\begin{align}
    \label{sminrBf3_subeq1}
    & {\bf\bar w}_{k_\text{SMINR}} = \argmax_{{\bf\bar w}_k}
    \frac{ ({\bf\bar w}_k {\bf\tilde h}_k d \sqrt{E_g})^2 -
    \sum_{\substack{j=1\\j\ne k}}^K ({\bf\bar w}_k {\bf\tilde h}_j s_j(L_j))^2}
    {\frac{\sigma_z^2}{2} } \\
    \label{sminrBf3_subeq2}
    &\text{subject to} \quad \|{\bf\bar w}_k\|_2^2 = 1.
\end{align}
\end{subequations}
Either by using Lagrange multipliers method or by rewriting (\ref{sminrBf3_eq}) as the
Rayleigh quotient problem
\begin{equation}\label{sminrRayleigh_eq}
    \max_{{\bf\bar w}_k}
    \frac{ {\bf\bar w}_k \left(d^2 E_g {\bf\tilde h}_k {\bf\tilde h}_k^T -
    \sum_{\substack{j=1\\j\ne k}}^K s_j^2(L_j) {\bf\tilde h}_j
    {\bf\tilde h}_j^T \right) {\bf\bar w}_k^T}
    {\frac{\sigma_z^2}{2} {\bf\bar w}_k {\bf\bar w}_k^T } ,
\end{equation}
the solution is given by
\begin{equation}
    {\bf\bar w}_{k_\text{SMINR}} = {\bf v}^T_\text{max}
    \left( d^2 E_g {\bf\tilde h}_k {\bf\tilde h}_k^T -
    \sum_{\substack{j=1\\j\ne k}}^K s_j^2(L_j) {\bf\tilde h}_j {\bf\tilde h}_j^T \right),
\end{equation}
i.e., the normalized eigenvector corresponding to the maximum eigenvalue of
$d^2 E_g {\bf\tilde h}_k \allowbreak {\bf\tilde h}_k^T -
\sum_{\substack{j=1\\j\ne k}}^K s_j^2(L_j) {\bf\tilde h}_j {\bf\tilde h}_j^T$.
Finally, ${\bf\ w}_{k_\text{SMINR}}$ can be obtained by
${\mathcal{T}_1^{-1}}({\bf\bar w}_{k_\text{SMINR}})$,
where the bijection ${\mathcal{T}_1}$ was defined in (\ref{transformations_eq}).

\section{Numerical Results}\label{secResults_uplink}
We consider a multiuser multiple access channel (MAC) with four single-antenna users,
each sending 8-PAM signals to a 4-antenna receiver simultaneously and at the
same carrier frequency.
The channel gains are assumed to be quasi static and follow a
Rayleigh distribution with unit variance. In other words, each element of the channel
is generated as a zero-mean and unit-variance i.i.d.\ CSCG random variable.
Since our focus is on the performance of receive beamforming methods rather
than on the effects of channel estimation, we assume that perfect CSI of all channels is
available at the receiver \cite{Peel05,Sadek07WC}.
At the receiver, i.i.d.\ Gaussian noise is added to the received signal.
All simulations are performed over 10,000 different
channel realizations and at
each channel realization a block of 1,000 symbols is transmitted from each user.
The above parameters are used in the following simulations unless stated otherwise.

\begin{figure}[tp]
  \centering
  \includegraphics[width=3.49in]{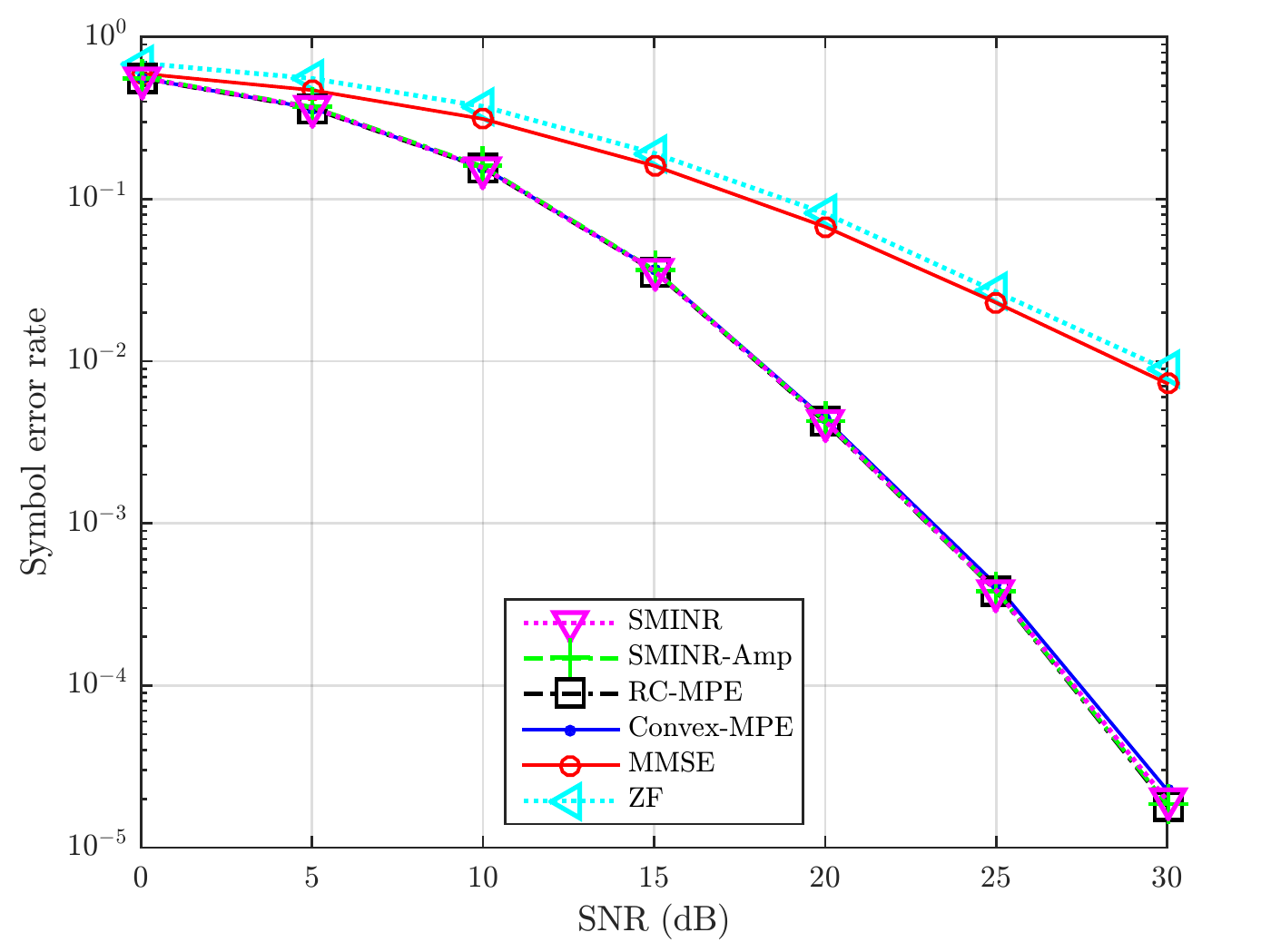}\\ 
  \caption{Average symbol error rates of users for $N=4$ receive antennas and $K=4$ users
  with 8-PAM modulation.}
  \label{fig_sminrPe}
\end{figure}
Fig. \ref{fig_sminrPe} compares the average symbol error rates of classical
ZF and MMSE receive beamforming with the proposed MPE, reduced-complexity MPE (RC-MPE),
amplitude version of maximum SMINR (SMINR-Amp), and heuristic maximum SMINR (SMINR) beamforming.
As expected, all the proposed methods substantially outperform ZF and MMSE beamforming.
For example, at a symbol error rate of $2.3 \times 10^{-2}$, all the
proposed beamforming methods
show a gain of about 9 dB compared to that of ZF and MMSE beamforming.
It is interesting to observe that average error probability of users
is nearly the same for all the proposed receive beamforming methods,
at all SNRs. Based on Proposition \ref{propositionSufficient},
it is expected that convex MPE beamforming
has the same performance as its reduced-complexity version, as confirmed in
Fig. \ref{fig_sminrPe}. However, it was not expected that the amplitude and
heuristic versions of maximum SMINR
perform as well as convex MPE beamforming. We expected
these beamforming methods to perform slightly worse than convex MPE and RC-MPE beamforming,
since they are designed based on minimization of an upper bound on the error probability.
However, as can be seen in Fig. \ref{fig_sminrPe}, maximum SMINR-Amp, at significantly
lower complexity, shows nearly the same performance as that of convex MPE beamforming.
This indicates that minimizing the
proposed upper bound on the error probability of each user
closely approximates minimization of the error probability function, at least
in this example.
Moreover, the near identical performance of the heuristic maximum
SMINR beamforming to that of convex MPE beamforming also indicates that
the SMINR function defined in (\ref{sminrPower_eq})
is an accurate reflection of the error probability function, again at least
for this example.

\begin{figure}[tp]
  \centering
  \includegraphics[width=3.49in]{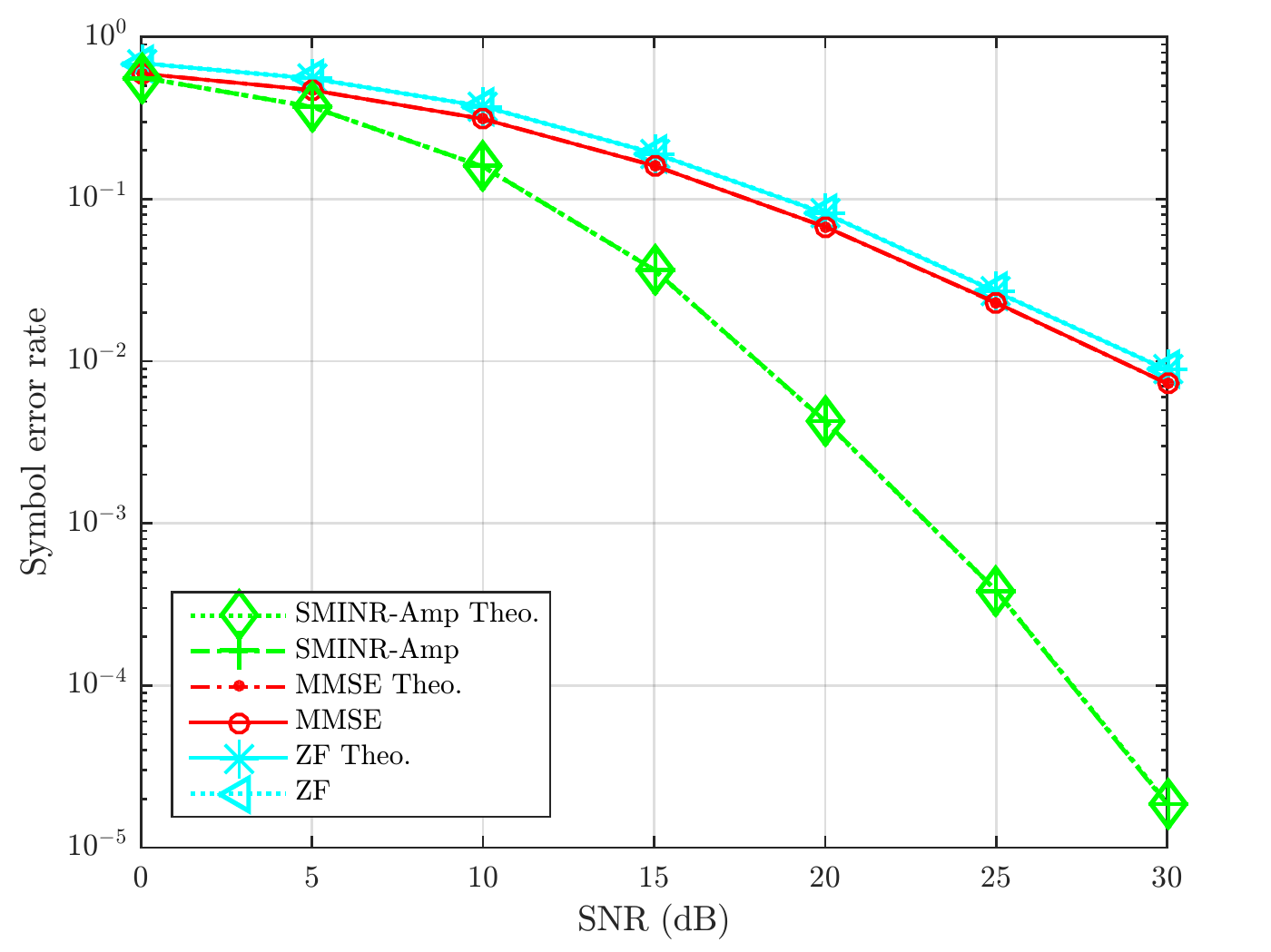}\\ 
  \caption{Average symbol error rates and theoretical symbol error rates of users
  for $N=4$ receive antennas and $K=4$ users with 8-PAM modulation.}
  \label{fig_sminrPeTheory}
\end{figure}
Fig. \ref{fig_sminrPeTheory} compares the average error probability
of users using Monte Carlo simulation and their theoretical counterparts
obtained by analytical expressions.
Analytically calculated BER curves in this figure are obtained by
substituting the calculated beamforming weights of users
for each channel realization into the error probability function
obtained in (\ref{userError_eq}).
As can be seen the calculated theoretical error
probability precisely predicts the error performance of users.

\begin{figure}[tp]
  \centering
  \includegraphics[width=3.49in]{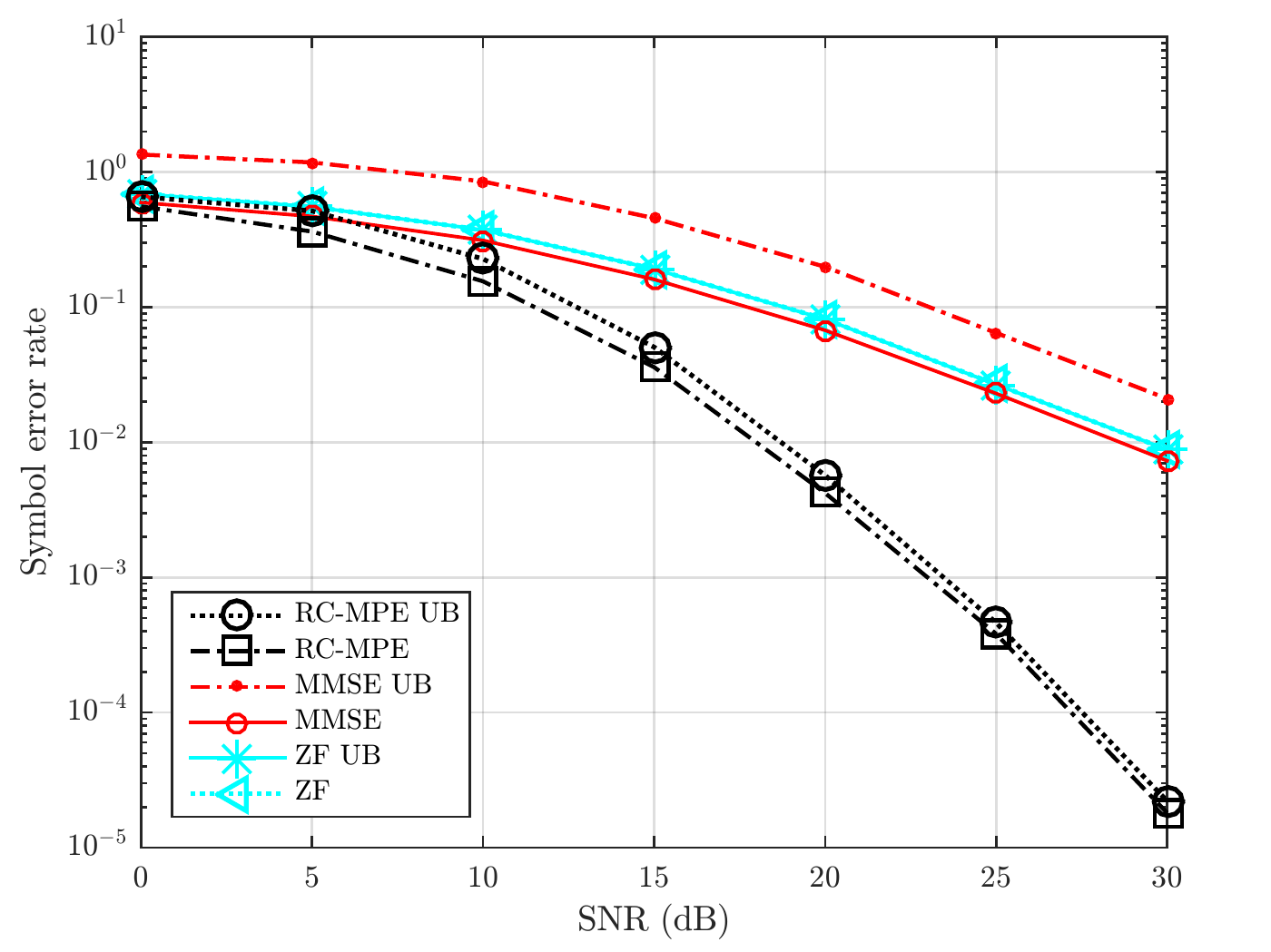}\\ 
  \caption{Symbol error rates and upper bounds on symbol error rates
  of users for $N=4$ receive antennas and $K=4$ users with 8-PAM modulation.}
  \label{fig_sminrPeUpperBound}
\end{figure}
Fig. \ref{fig_sminrPeUpperBound} compares the average error probability
of users and their corresponding upper bounds given by (\ref{userErrorUpperBound_eq}).
As can be seen, the upper bound curves are either above the
error probability curves as in case of MMSE and RC-MPE beamforming or lie on top
of the error probability curves as in case of ZF beamforming.
In zero-forcing, the upper bound lies exactly over the error probability curve.
In other words, in ZF beamforming the proposed upper bound on the error probability
is equal to the exact error probability.
The zero-forcing beamformer enforces the beamforming weight vector of a user to be orthogonal to
the channels of other users, i.e., ${\bf w}_k {\bf h}_j=0,~k \ne j$.
Therefore, in ZF beamforming the error probability (\ref{userError_eq})
and the upper bound on the error probability (\ref{userErrorUpperBound_eq}) are equivalent.
In Fig. \ref{fig_sminrPeUpperBound}, it can also be seen that at low SNRs the upper bound
on error probability of MMSE beamforming is greater than one. This indicates that
for the beamforming weights obtained by MMSE beamforming, the argument of the
$Q$-function in (\ref{userErrorUpperBound_eq}) is not always greater than zero.
It can also be seen that in reduced-complexity MPE beamforming,
the upper bound closely approximates the error probability.
Overall, based on Fig. \ref{fig_sminrPeUpperBound}, it is inferred that
the tightness of the proposed upper bound is not the same for different beamforming
techniques but depends on the values of the beamforming weight vectors of the users.

\begin{figure}[tp]
  \centering
  \includegraphics[width=3.49in]{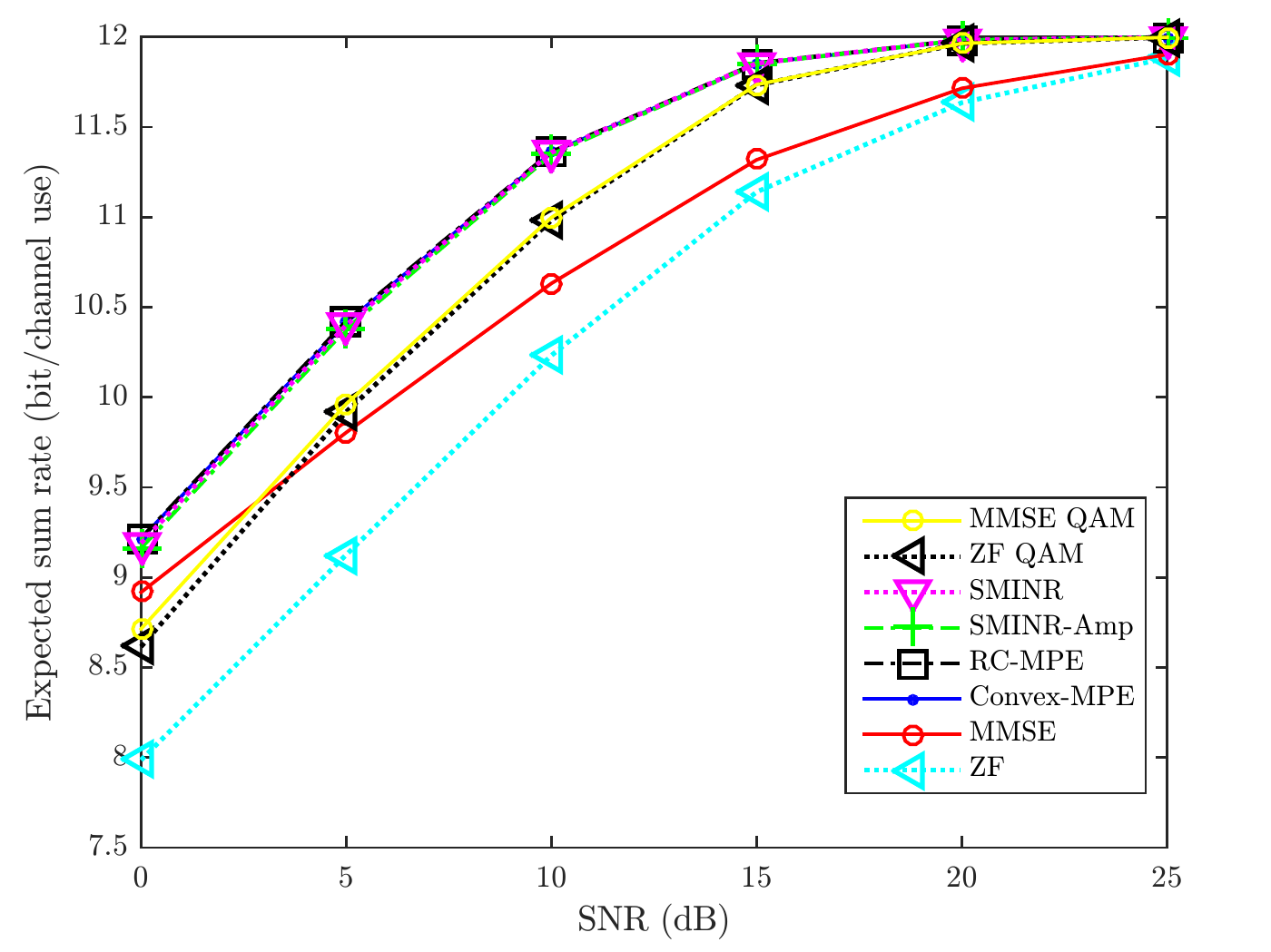}\\ 
  \caption{Average sum rates when $N=4$ and $K=2$ with 64-QAM modulated signals
  for ZF-QAM and MMSE-QAM and when $N=4$ and $K=4$ with 8-PAM modulated signals for
  the other simulated beamforming methods.}
  \label{fig_sminrRateBit}
\end{figure}
So far, we have compared the proposed beamforming techniques for 1D
signalling with classical beamforming using 1D signalling. It would also
be instructive to extend the above comparison to two-dimensionally modulated
signals.
Fig. \ref{fig_sminrRateBit} compares the expected sum rate (throughput) of users
employing the proposed beamforming methods of one-dimensionally modulated
signals and classical beamforming of both one-dimensionally and two-dimensionally
modulated signals.
As can be seen, at all SNRs, the proposed convex-MPE, RC-MPE, maximum SMINR-Amp, and maximum SMINR
beamforming methods achieve higher sum rates than ZF and MMSE beamforming. It should be
remarked that in addition to the sum rate of four users with 8-PAM modulation and
the proposed beamforming methods,
the sum rate of two users with 64-QAM modulation using ZF and MMSE receive beamforming are also
included in Fig. \ref{fig_sminrRateBit}. Theoretically, four users with 8-PAM signalling as well as
two users with 64-QAM signalling achieve a maximum bit rate of 12 bits/channel use.
Therefore, it is interesting to observe that the proposed beamforming methods which are
developed for one-dimensionally modulated signals not only outperform classical beamforming
of one-dimensionally modulated signals (as expected), but also they outperform classical
beamforming of their counterpart two-dimensional modulations.

\begin{figure}[tp]
  \centering
  \includegraphics[width=3.49in]{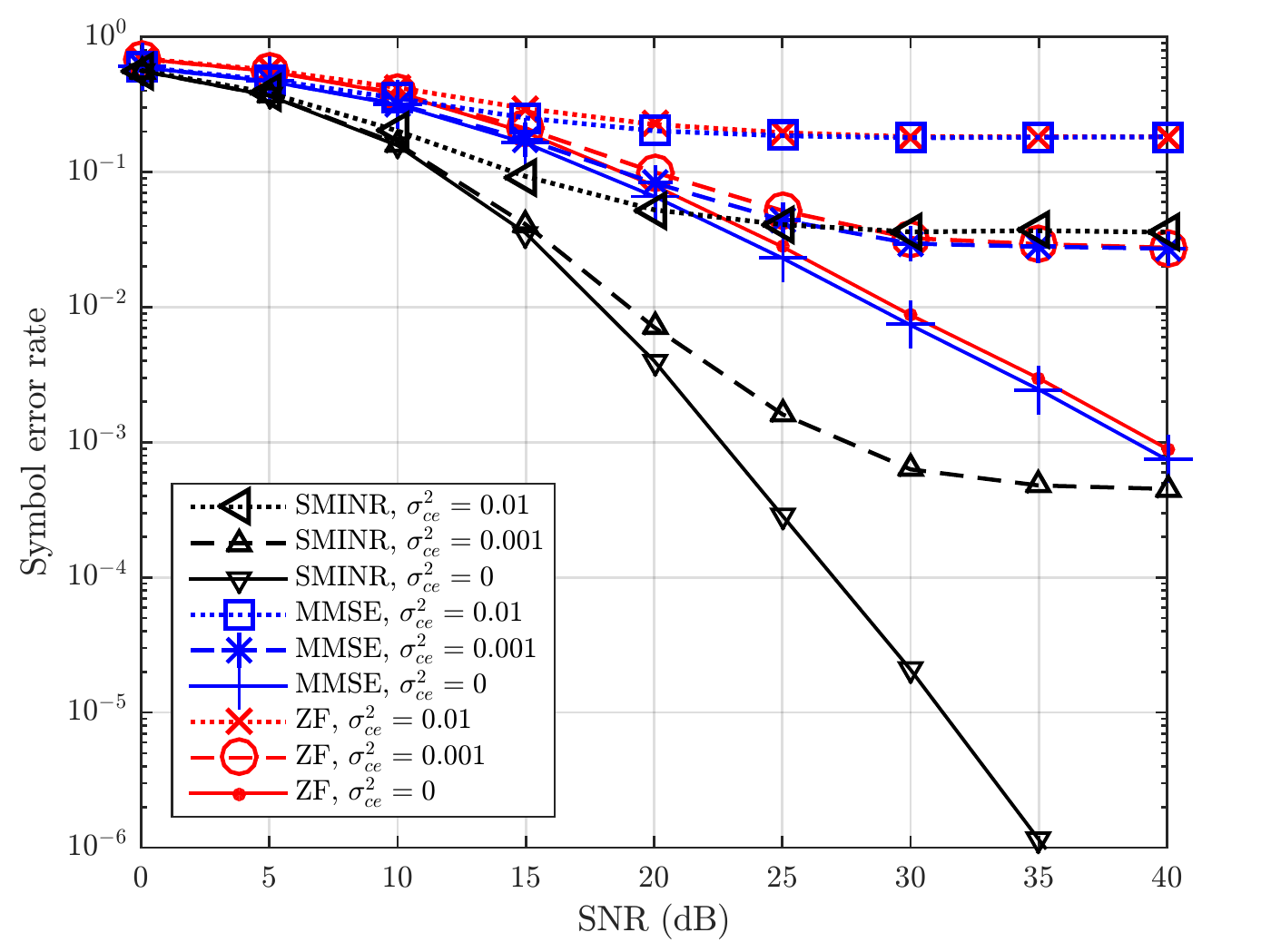}\\ 
  \caption{Average symbol error rates of users for $N=4$ receive antennas and $K=4$ users
  with 8-PAM modulation assuming imperfect CSI.}
  \label{fig_sminrPeImperfectChannel}
\end{figure}
As has been seen so far,
power-based maximum SMINR beamforming with closed-form solution exhibits
superior performance compared to conventional ZF and MMSE beamforming.
Therefore, it would be interesting to compare its sensitivity to
imperfect CSI with that of ZF and MMSE beamforming.
In Fig. \ref{fig_sminrPeImperfectChannel}, average symbol error rates
of users assuming both perfect and imperfect CSI are presented for maximum
SMINR, ZF, and MMSE beamforming. For imperfect CSI, it is assumed that
channel estimation error is normally distributed with zero mean and variance of
either $\sigma_{ce}^2 = 0.01$ or $0.001$.
As mentioned earlier, the channel gains have a CSCG distribution with
zero mean and unit variance\footnote{Considering unit variance for the channel gains,
$\sigma_{ce}^2 = 0.01,~ 0.001$ correspond to channel gain estimation
SNRs of 20 and 30 dB, respectively.}.
It is seen in Fig. \ref{fig_sminrPeImperfectChannel} that
as estimation error increases, the error probability increases.
In this figure and also in Fig. \ref{fig_sminrPe},
it can be seen that the proposed maximum SMINR beamforming with perfect CSI
outperforms ZF and MMSE beamforming methods with perfect CSI.
In Fig. \ref{fig_sminrPeImperfectChannel}, it is observed that
the proposed maximum SMINR beamforming with imperfect CSI also
outperforms classical beamforming methods with imperfect CSI. For example,
from Fig. \ref{fig_sminrPeImperfectChannel},
at SNR of 40 dB and channel error variance of $\sigma_{ce}^2 = 0.001$, bit error error
rate of maximum SMINR beamforming is $4.5 \times 10^{-4}$, while bit error
rates of ZF and MMSE beamforming are about $2.7 \times 10^{-2}$.
It is interesting to note that up to SNR of 40 dB,
the maximum SMINR with $\sigma_{ce}^2 = 0.001$ not only
outperforms both ZF and MMSE with $\sigma_{ce}^2 = 0.001$ but also outperforms
ZF and MMSE with perfect CSI.

\section{Conclusion}\label{secConclusion_uplink}
In this paper, it has been shown that by exploiting the type of modulation in the design
of receive beamforming, the performance of a multiuser multiple-access communications
system can be tremendously improved.
The error probability of each user can be calculated and minimized to obtain the optimum
beamforming weights. This highly nonlinear optimization problem was transformed
to a convex optimization problem and two reduced-complexity
versions of the problem were also introduced and solved numerically. Finally, the error
probability in a multiuser scenario resulted in the development of a new metric
called signal minus interference to noise ratio, where its maximization
resulted in a closed-form solution for receive beamforming weights based
on a simple eigenvalue decomposition. It has been shown that all the proposed beamforming
techniques outperform the classical zero-forcing and MMSE beamforming.

\appendices
\section{Proof of Proposition
\ref{propositionPositiveArgQ}}\label{subSecAppendPropositionPositiveArgQ}
Assume that there exists a $b=b_1$ such that
\begin{equation}\label{argQLess0_eq}
    \Re\{{\bf w}_k {\bf h}_k d \sqrt{E_g} -
    {\bf w}_k {\bf H}_{\bar k} {\bf s}_{\bar k}(b_1)\} < 0.
\end{equation}
Therefore,
\begin{equation}
    \frac{ \Re\{{\bf w}_k {\bf h}_k d \sqrt{E_g} -
    {\bf w}_k {\bf H}_{\bar k} {\bf s}_{\bar k}(b_1)\}} {\frac{\sigma_z}{\sqrt{2}} \|{\bf w}_k\|_2}
    < 0.
\end{equation}
Hence,
\begin{equation}
    Q\left(\frac{ \Re\{{\bf w}_k {\bf h}_k d \sqrt{E_g} -
    {\bf w}_k {\bf H}_{\bar k} {\bf s}_{\bar k}(b_1)\}} {\frac{\sigma_z}{\sqrt{2}} \|{\bf w}_k\|_2}
    \right) > \frac{1}{2}.
\end{equation}
Consequently, error probability is written as
\begin{align}
    \nonumber
    &P_{e_k} = \frac{L_k-1}{N_b} \\
    &+\frac{2(L_k-1)}{N_b} \sum_{\substack{b=1\\b \ne b_1}}^{N_{p_k}}
    Q\left(\frac{ \Re\{{\bf w}_k {\bf h}_k d \sqrt{E_g} -
    {\bf w}_k {\bf H}_{\bar k} {\bf s}_{\bar k}(b)\}}
    {\frac{\sigma_z}{\sqrt{2}} \|{\bf w}_k\|_2}\right).
\end{align}
If $d\sqrt{E_g}$ approaches infinity, error probability of user $k$ approaches
$\lim_{d\sqrt{E_g} \rightarrow +\infty} P_{e_k} = \frac{L_k-1}{N_b}$.
In other words, there always exists an error floor of $\frac{L_k-1}{KN_b}=\frac{L_k-1}{L_k N_p K}$,
because the number of users is limited and consequently is $N_p$.

\section{Proof of Property \ref{propertyScaleQ}}\label{subSecAppendPropertyScaleQ}
Let ${\bf w}^{\prime}_k=c{\bf w}_k$, where $c \in \mathbb{R} ^+$. We have
\begin{align}
    \nonumber
    &P_{e_k}({\bf w}^{\prime}_k) \\
    \nonumber
    &=\frac{2(L_k-1)}{N_b} \sum_{b=1}^{N_{p_k}}
    Q\left(\frac{ \Re\{{\bf w}'_k {\bf h}_k d \sqrt{E_g} -
    {\bf w}'_k {\bf H}_{\bar k} {\bf s}_{\bar k}(b)\}}
    {\frac{\sigma_z}{\sqrt{2}} \|{\bf w}'_k\|_2}\right)\\
    \nonumber
    &=\frac{2(L_k-1)}{N_b} \sum_{b=1}^{N_{p_k}}
    Q\!\left(\!\frac{ \Re\{c{\bf w}_k {\bf h}_k d \sqrt{E_g} \!-\!
    c{\bf w}_k {\bf H}_{\bar k} {\bf s}_{\bar k}(b)\}}
    {\frac{\sigma_z}{\sqrt{2}} \|c{\bf w}_k\|_2}\!\right)\\
    &= P_{e_k}({\bf w}_k).
\end{align}

\section{Proof of Theorem
\ref{theoremUniqueGlobalMinimizer}}\label{subSecAppendTheoremUniqueGlobalMinimizer}
The minimization problem of error probability of user $k$
is considered over the following feasible set:
\begin{align}\label{feasibleSet_eq}
    \nonumber
    {\mathcal F}_k = &
    \{ {\bf w}_k | {\bf w}_k {\bf w}_k^H = 1  \wedge ~\\
    &\Re\{{\bf w}_k {\bf h}_k d \sqrt{E_g} -
    {\bf w}_k {\bf H}_{\bar k} {\bf s}_{\bar k}(b)\} \ge 0, ~ 1 \leq b \leq N_{p_k} \} .
\end{align}
Assume that ${\bf w}_{k_1} \in {\mathcal F}_k$ is a global minimizer of the
optimization problem (\ref{mpeConstrained_eq}), and ${\bf w}_{k_2} \in {\mathcal F}_k$ is
a local minimizer of the problem such that
\begin{equation}\label{localMinInequality_eq}
    P_{e_k}({\bf w}_{k_1}) < P_{e_k}({\bf w}_{k_2}) .
\end{equation}
Assuming $0 < \alpha < 1$, we define ${\bf w}_{k_0}$ as
\begin{equation}
    {\bf w}_{k_0} = \frac {\alpha {\bf w}_{k_1}+(1-\alpha){\bf
    w}_{k_2}} {\| \alpha {\bf w}_{k_1}+(1-\alpha){\bf w}_{k_2} \|_2} .
\end{equation}
Therefore, we have $\|{\bf w}_{k_0}\|=1$, and for $1 \leq b \leq
N_{p_k}$, we have $\Re\{{\bf w}_{k_0} {\bf h}_k d \sqrt{E_g} -
{\bf w}_{k_0} {\bf H}_{\bar k} {\bf s}_{\bar k}(b)\} \ge 0$.
Hence, it can be inferred that ${\bf w}_{k_0} \in {\mathcal F}_k$.
It is also obvious that
\begin{equation}
    \| \alpha {\bf w}_{k_1}+(1-\alpha){\bf w}_{k_2} \|_2 \leq \alpha
    \|{\bf w}_{k_1}\|_2 + (1-\alpha)\|{\bf w}_{k_2}\|_2 = 1 .
\end{equation}
Consequently,
\begin{align}\label{qFuncNumeratorInequality_eq}
    \nonumber
    &\Re\{{\bf w}_{k_0} {\bf h}_k d \sqrt{E_g} -
    {\bf w}_{k_0} {\bf H}_{\bar k} {\bf s}_{\bar k}(b)\} \\
    \nonumber
    &\ge
    \alpha \Re\{{\bf w}_{k_1} {\bf h}_k d \sqrt{E_g} -
    {\bf w}_{k_1} {\bf H}_{\bar k} {\bf s}_{\bar k}(b)\} \\
    &+(1-\alpha)\Re\{{\bf w}_{k_2} {\bf h}_k d \sqrt{E_g} -
    {\bf w}_{k_2} {\bf H}_{\bar k} {\bf s}_{\bar k}(b)\},~
    1 \leq b \leq N_{p_k}.
\end{align}
Therefore, we have 
\begin{align}\label{qFuncInequality_eq}
    \nonumber
    & Q\left( \frac{ \Re\{{\bf w}_{k_0} {\bf h}_k d \sqrt{E_g} -
    {\bf w}_{k_0} {\bf H}_{\bar k} {\bf s}_{\bar k}(b)\}}
    {\frac{\sigma_z}{\sqrt{2}}} \right) \\
    \nonumber
    &\leq Q\left( \frac{ \alpha \Re\{{\bf w}_{k_1} {\bf h}_k d \sqrt{E_g} -
    {\bf w}_{k_1} {\bf H}_{\bar k} {\bf s}_{\bar k}(b)\} +
    (1-\alpha) \Re\{{\bf w}_{k_2} {\bf h}_k d \sqrt{E_g} -
    {\bf w}_{k_2} {\bf H}_{\bar k} {\bf s}_{\bar k}(b)\}}
    {\frac{\sigma_z}{\sqrt{2}}}
    \right) \\
    \nonumber
    &\leq \alpha Q\left( \frac{\Re\{{\bf w}_{k_1} {\bf h}_k d \sqrt{E_g} -
    {\bf w}_{k_1} {\bf H}_{\bar k} {\bf s}_{\bar k}(b)\}}
    {\frac{\sigma_z}{\sqrt{2}}} \right) \\
    &+ (1 - \alpha) Q\left( \frac{\Re\{{\bf w}_{k_2} {\bf h}_k d \sqrt{E_g} -
    {\bf w}_{k_2} {\bf H}_{\bar k} {\bf s}_{\bar k}(b)\}}
    {\frac{\sigma_z}{\sqrt{2}}} \right), \quad 1 \leq b \leq N_{p_k},
\end{align}
where the first inequality results from (\ref{qFuncNumeratorInequality_eq}) and due
to the fact that $Q(x)$ is a decreasing function for $x \geq
0$, and the second inequality stands because $Q(x)$ is a
convex function for $x \geq 0$. 

From (\ref{userError_eq}) and (\ref{qFuncInequality_eq}), it can be inferred that
\begin{align}\label{eq22}
    \nonumber
    &P_{e_k}({\bf w}_{k_0}) \\
    \nonumber
    &= \frac{2(L_k-1)}{N_b} \sum_{b=1}^{N_{p_k}}
    Q\left( \frac{\Re\{{\bf w}_{k_0} {\bf h}_k d \sqrt{E_g} -
    {\bf w}_{k_0} {\bf H}_{\bar k} {\bf s}_{\bar k}(b)\}}
    {\frac{\sigma_z}{\sqrt{2}} \|{\bf w}_{k_0}\|_2} \right) \\
    \nonumber
    &\le \frac{\alpha 2(L_k-1)}{N_b} \sum_{b=1}^{N_{p_k}}
    Q\left( \frac{\Re\{{\bf w}_{k_1} {\bf h}_k d \sqrt{E_g} -
    {\bf w}_{k_1} {\bf H}_{\bar k} {\bf s}_{\bar k}(b)\}}
    {\frac{\sigma_z}{\sqrt{2}} \|{\bf w}_{k_1}\|_2} \right) \\
    \nonumber
    &+\! \frac{(1 \!-\! \alpha)2(L_k \!-\! 1)}{N_b} \!\sum_{b=1}^{N_{p_k}} \!\!
    Q\!\left(\! \frac{\Re\{\!{\bf w}_{k_2} {\bf h}_k d \sqrt{E_g} \!-\!
    {\bf w}_{k_2} {\bf H}_{\bar k} {\bf s}_{\bar k}(b)\!\}}
    {\frac{\sigma_z}{\sqrt{2}} \|{\bf w}_{k_2}\|_2} \!\right) \\
    &=\! \alpha P_{e_k}({\bf w}_{k_1}) \!+\! (1 \!-\! \alpha) P_{e_k}({\bf
    w}_{k_2}) \!<\! P_{e_k}({\bf w}_{k_2}) ,~ \forall \alpha \!\in\! (0,1),
\end{align}
where the last inequality is due to the assumption of the proof,
i.e., ${\bf w}_{k_1}$ is the global minimizer of $P_{e_k}$. Now, let $\alpha
\rightarrow 0$, ${\bf w}_{k_0} \rightarrow {\bf
w}_{k_2}$. Hence, in a small neighborhood of ${\bf w}_{k_2}$,
there always exists a ${\bf w}_{k_0}$, so that $P_{e_k}({\bf
w}_{k_0}) < P_{e_k}({\bf w}_{k_2})$, i.e., ${\bf w}_{k_2}$ is not
a local minimizer. In other words, there does not exist any local
minimizer such that (\ref{localMinInequality_eq}) holds. Therefore, it can be
concluded that either no local minimizer exists, which
proves the theorem, or there exists a local minimizer such that
$P_{e_k}({\bf w}_{k_1}) \geq P_{e_k}({\bf w}_{k_2})$. However,
since ${\bf w}_{k_1}$ is a global minimizer of $P_{e_k}({\bf
w}_k)$, we have $P_{e_k}({\bf w}_{k_1}) \leq P_{e_k}({\bf
w}_{k_2})$. Therefore, it can be concluded that $P_{e_k}({\bf
w}_{k_1}) = P_{e_k}({\bf w}_{k_2})$, i.e., the local minimizer (if
exists) is also a global minimizer.

To show the uniqueness of the global minimizer, first we consider
the following set:
\begin{align}
    \nonumber
    {\mathcal F}_k^0 =  &\{ {\bf w}_k | {\bf w}_k {\bf w}_k^H = 1  \wedge ~ \\
    &\Re\{{\bf w}_k {\bf h}_k d \sqrt{E_g} -
    {\bf w}_k {\bf H}_{\bar k} {\bf s}_{\bar k}(b)\} = 0, ~ 1 \leq b \leq N_{p_k} \} .
\end{align}
It is obvious that each point in this set is a global maximizer of
error probability function in (\ref{userError_eq}) constrained by the set
defined in (\ref{feasibleSet_eq}), because the arguments of
all $Q$-functions in error probability (\ref{userError_eq}) will be zero. Therefore, to
solve the minimization problem it is sufficient to solve the
problem over the set ${\mathcal F}_k^1={\mathcal F}_k - {\mathcal F}_k^0$.
The error probability of user $k$, $P_{e_k}({\bf w}_k)$, is
strictly convex on ${\mathcal F}_k^1$, because $Q(x)$ is strictly
convex for $x > 0$. Assume that ${\bf w}_{k_1} \neq {\bf w}_{k_2}$
are two global minimizers of the optimization problem
(\ref{mpeConstrained_eq}). We define ${\bf w}_{k_0}$ as follows:
\begin{equation}
    {\bf w}_{k_0} = \frac {\alpha {\bf w}_{k_1}+(1 - \alpha){\bf
    w}_{k_2}} {\| \alpha {\bf w}_{k_1}+(1 - \alpha){\bf w}_{k_2} \|_2},
    \quad \forall \alpha \in (0,1).
\end{equation}
Since ${\bf w}_{k_1}$ is a global minimizer, it is obvious that
\begin{equation}\label{globalMinInequality1_eq}
    P_{e_k}({\bf w}_{k_0}) \geq P_{e_k}({\bf w}_{k_1}) .
\end{equation}
On the other hand, we have
\begin{equation}\label{globalMinInequality2_eq}
    P_{e_k}({\bf w}_{k_0}) < \alpha P_{e_k}({\bf w}_{k_1}) + (1 -
    \alpha) P_{e_k}({\bf w}_{k_2}) = P_{e_k}({\bf w}_{k_1}) ,
\end{equation}
because $P_{e_k}({\bf w}_k)$ is strictly convex on ${\mathcal F}_k^1$. Since
(\ref{globalMinInequality2_eq}) contradicts (\ref{globalMinInequality1_eq}),
it can be inferred that the global minimizer is unique.

\section{Proof of Proposition
\ref{propositionSufficient}}\label{subSecAppendPropositionSufficient}
%

To prove this proposition we first prove the sufficient condition by showing
that the left hand side (LHS) of
(\ref{propositionSufficient_eq}) is a lower bound on the LHS of the inequality
(\ref{mpeConstrained3_subeq3}) for all $b$.

For $1 \le j, k \le K$, we have
\begin{align}
    \nonumber
    &|\Re\{{\bf w}_k {\bf h}_j s_j(L_j)\}| = |\Re\{{\bf w}_k {\bf h}_j\}|.|s_j(L_j)| \\
    \nonumber
    &\ge |\Re\{{\bf w}_k {\bf h}_j\}| . |s_j(l_j)|
    \ge \Re\{{\bf w}_k {\bf h}_j\} s_j(l_j) \\
    &= \Re\{{\bf w}_k {\bf h}_j s_j(l_j) \} , \quad \forall l_j \in \{1,\cdots,L_j\}.
\end{align}
Therefore,
\begin{align}
    &\sum_{\substack{j=1\\j \ne k}} |\Re\{{\bf w}_k {\bf h}_j s_j(L_j)\}| \ge
    \sum_{\substack{j=1\\j \ne k}} \Re\{{\bf w}_k {\bf h}_j s_j(l_j) \} \\
    \nonumber
    &= \Re\{{\bf w}_k {\bf H}_{\bar k} {\bf s}_{\bar k}(b)\}, \quad 1\le b \le N_{p_k}.
\end{align}
Hence,
\begin{align}\label{qFuncArgLowerBound_eq}
    \nonumber
    & \Re\{{\bf w}_k {\bf h}_k d \sqrt{E_g} \} - \sum_{\substack{j=1\\j\ne k}}^K
    |\Re\{{\bf w}_k {\bf h}_j s_j(L_j)\}| \\
    \nonumber
    & \le \Re\{{\bf w}_k {\bf h}_k d \sqrt{E_g} \} -
    \Re\{{\bf w}_k {\bf H}_{\bar k} {\bf s}_{\bar k}(b)\} \\
    &= \Re\{{\bf w}_k {\bf h}_k d \sqrt{E_g} -
    {\bf w}_k {\bf H}_{\bar k} {\bf s}_{\bar k}(b)\}, \quad 1\le b \le N_{p_k}.
\end{align}
Thus, if $\Re\{ {\bf w}_k {\bf h}_k d\sqrt{E_g} \} - \sum_{\substack{j=1 \\ j \ne k}}^K
|\Re\{ {\bf w}_k {\bf h}_j s_j(L_j) \}| \ge 0$ then
$\Re\{ {\bf w}_k {\bf h}_k d\sqrt{E_g}  -
{\bf w}_k {\bf H}_{\bar k} \allowbreak {\bf s}_{\bar k}(b) \} \ge 0,~1 \le b \le N_{p_k}$.

To prove the necessary condition, we use contradiction. Let us assume that
$\Re\{ {\bf w}_k {\bf h}_k d\sqrt{E_g} \} - \sum_{\substack{j=1 \\ j \ne k}}^K
|\Re\{ {\bf w}_k {\bf h}_j s_j(L_j) \}| < 0$. Therefore, we have
\begin{align}
    \nonumber
    &\Re\{ {\bf w}_k {\bf h}_k d\sqrt{E_g} \} - \sum_{\substack{j=1 \\ j \ne k}}^K
    |\Re\{ {\bf w}_k {\bf h}_j s_j(L_j) \}|  \\
    \nonumber
    &= \Re\{ {\bf w}_k {\bf h}_k d\sqrt{E_g} \} - \sum_{\substack{j=1 \\ j \ne k}}^K
    \Re\{ {\bf w}_k {\bf h}_j\} \sgn(\Re\{ {\bf w}_k {\bf h}_j\}) s_j(L_j) \\
    \nonumber
    &= \Re\{ {\bf w}_k {\bf h}_k d\sqrt{E_g} \} - \sum_{\substack{j=1 \\ j \ne k}}^K
    \Re\{ {\bf w}_k {\bf h}_j\} s_j(\ell_j)  \\
    &= \Re\{ {\bf w}_k {\bf h}_k d\sqrt{E_g} \} -
    \Re\{ {\bf w}_k {\bf H}_{\bar k} {\bf s}_{\bar k}(\beta)\} < 0,
\end{align}
where ${\bf s}_{\bar k}(\beta) = [s_1(\ell_1), \cdots, s_{k-1}(\ell_{k-1}), s_{k+1}(\ell_{k+1}),
\cdots, \allowbreak s_K(\ell_K)]^T$, and for $j\ne k$,
$\ell_j = L_j$ if $\sgn(\Re\{ {\bf w}_k {\bf h}_j\}) =1$ and
$\ell_j = 1$ if $\sgn(\Re\{ {\bf w}_k {\bf h}_j\})=-1$.
Therefore, there exists a $b=\beta$ such that $\Re\{ {\bf w}_k {\bf h}_k d\sqrt{E_g} \} \allowbreak -
\Re\{ {\bf w}_k {\bf H}_{\bar k} {\bf s}_{\bar k}(\beta)\}  < 0$, i.e., at least one constraint
in (\ref{mpeConstrained3_subeq3}) is not satisfied.

\section{Proof of Claim \ref{claimConvexSet}}\label{subSecAppendClaimConvexSet}
Using the definition of a convex set \cite{BoydBook04}, it is assumed
that ${\bf w}_{k_1}$ and ${\bf w}_{k_2}$ are two arbitrary points
in the set defined by (\ref{mpeReduced_subeq3}), i.e.,
\begin{equation}\label{reducedSet_eq}
    \{{\bf w}_k | \Re\{{\bf w}_k {\bf h}_k d \sqrt{E_g} \} -
    \sum_{\substack{j=1\\j\ne k}}^K \left|\Re\{{\bf w}_k {\bf h}_j s_j(L_j)\}\right| \ge 0\}.
\end{equation}
For any $0 \le \alpha \le 1$,
\begin{align}
    \nonumber
    & |\Re\{(\alpha{\bf w}_{k_1}+(1-\alpha){\bf w}_{k_2}) {\bf h}_j s_j(L_j)\}| \\
    \nonumber
    & = |\alpha\Re\{{\bf w}_{k_1} {\bf h}_j s_j(L_j)\} +
    (1-\alpha)\Re\{{\bf w}_{k_2} {\bf h}_j s_j(L_j)\}| \\
    \nonumber
    & \le |\alpha\Re\{{\bf w}_{k_1} {\bf h}_j s_j(L_j)\}| +
    |(1-\alpha)\Re\{{\bf w}_{k_2} {\bf h}_j s_j(L_j)\}| \\
    & = \alpha |\Re\{{\bf w}_{k_1} {\bf h}_j s_j(L_j)\}| +
    (1-\alpha) |\Re\{{\bf w}_{k_2} {\bf h}_j s_j(L_j)\}|,
\end{align}
where the inequality is the result of the triangle inequality.
Therefore,
\begin{align}
    \nonumber
    & \sum_{\substack{j=1\\j\ne k}}^K
    |\Re\{(\alpha{\bf w}_{k_1}+(1-\alpha){\bf w}_{k_2}) {\bf h}_j s_j(L_j)\}| \\
    & \le\! \sum_{\substack{j=1\\j\ne k}}^K
    \alpha |\Re\{{\bf w}_{k_1} {\bf h}_j s_j(L_j)\}| \!+\!
    \sum_{\substack{j=1\\j\ne k}}^K
    (1 \!-\! \alpha) |\Re\{{\bf w}_{k_2} {\bf h}_j s_j(L_j)\}|,
\end{align}
and consequently,
\begin{align}
    \nonumber
    & \Re\{(\alpha{\bf w}_{k_1}+(1-\alpha){\bf w}_{k_2})
    {\bf h}_k d \sqrt{E_g} \} \\
    \nonumber
    &- \sum_{\substack{j=1\\j\ne k}}^K
    |\Re\{(\alpha{\bf w}_{k_1}+(1-\alpha){\bf w}_{k_2}) {\bf h}_j s_j(L_j)\}| \\
    \nonumber
    &\ge \alpha\left(\Re\{{\bf w}_{k_1} {\bf h}_k d \sqrt{E_g} \} -
    \sum_{\substack{j=1\\j\ne k}}^K
    |\Re\{{\bf w}_{k_1} {\bf h}_j s_j(L_j)\}| \right) \\
    \nonumber
    &+ (1-\alpha) \left(\Re\{{\bf w}_{k_2} {\bf h}_k d \sqrt{E_g} \} -
    \sum_{\substack{j=1\\j\ne k}}^K
    |\Re\{{\bf w}_{k_2} {\bf h}_j s_j(L_j)\}| \right) \\
    &\ge 0.
\end{align}
The last inequality holds because ${\bf w}_{k_1}$ and ${\bf w}_{k_2}$ are
in the set (\ref{reducedSet_eq}). Thus, (\ref{mpeReduced_subeq3}) is a
convex set.


\begin{thebibliography}{10}

\bibitem{Telatar99}
I.~E. Telatar, ``Capacity of multi-antenna {G}aussian channels,''
  \emph{European Trans. Telecommun.}, vol.~10, no.~6, pp. 585--595, Nov. 1999.

\bibitem{Paulraj04}
A.~J. Paulraj, D.~A. Gore, R.~U. Nabar, and H.~Bolcskei, ``An overview of
  {MIMO} communications - {A} key to gigabit wireless,'' \emph{Proc. {IEEE}},
  vol.~92, no.~2, pp. 198--218, Feb. 2004.

\bibitem{GesbertSPMag07}
D.~Gesbert, M.~Kountouris, R.~W. Heath~Jr., C.-B. Chae, and T.~S\"{a}lzer,
  ``From single-user to multiuser communications: Shifting the {MIMO}
  paradigm,'' \emph{{IEEE} Signal Process. Mag.}, vol.~24, no.~5, pp. 36--46,
  Sep. 2007.

\bibitem{WeiYu04}
W.~Yu and J.~M. Cioffi, ``Sum capacity of {G}aussian vector broadcast
  channels,'' \emph{{IEEE} Trans. Inf. Theory}, vol.~50, no.~9, pp. 1875--1892,
  Sep. 2004.

\bibitem{Lim13}
C.~Lim, T.~Yoo, B.~Clerckx., B.~Lee, and B.~Shim, ``Recent trend of multiuser
  {MIMO} in {LTE-A}dvanced,'' \emph{{IEEE} Commun. Mag.}, vol.~51, no.~3, pp.
  127--135, Mar. 2013.

\bibitem{Veen88}
B.~D.~V. Veen and K.~M. Buckley, ``Beamforming: {A} versatile approach to
  spatial filtering,'' \emph{Proc. {IEEE}}, vol.~5, pp. 4--24, Apr 1988.

\bibitem{LitvaBook96}
J.~Litva and T.~K.~Y. Lo, \emph{Digital Beamforming in Wireless
  Communications}.\hskip 1em plus 0.5em minus 0.4em\relax London, U.K.: Artech,
  1996.

\bibitem{Sidiropoulos06}
N.~D. Sidiropoulos, T.~N. Davidson, and Z.-Q. Luo, ``Transmit beamforming for
  physical-layer multicasting,'' \emph{{IEEE} Trans. Signal Process.}, vol.~54,
  no.~6, pp. 2239--2251, Jun. 2006.

\bibitem{Gershman10}
A.~B. Gershman, N.~D. Sidiropoulos, S.~Shahbazpanahi, M.~Bengtsson, and
  B.~Ottersten, ``Convex optimization-based beamforming,'' \emph{{IEEE} Signal
  Process. Mag.}, vol.~27, no.~3, pp. 62--75, May 2010.

\bibitem{MajidICT11}
M.~Bavand and P.~Azmi, ``Successive detection based minimum probability of
  error beamforming,'' in \emph{Proc. 18th {IEEE} Int. Conf. Telecommun.}, May
  2011, pp. 357--362.

\bibitem{GodaraPartI97}
L.~C. Godara, ``Applications of antenna arrays to mobile communications, {Part
  I:} performance improvement, feasibility, and system considerations,''
  \emph{Proc. {IEEE}}, vol.~85, no.~7, pp. 1031 --1060, Jul. 1997.

\bibitem{GodaraPartII97}
------, ``Application of antenna arrays to mobile communications, {Part II}:
  Beam-forming and direction-of-arrival considerations,'' \emph{Proc. {IEEE}},
  vol.~85, no.~8, pp. 1195--1245, Aug. 1997.

\bibitem{Zhou12}
L.~Liu, R.~Chen, S.~Geirhofer, K.~Sayana, Z.~Shi, and Y.~Zhou, ``Downlink
  {MIMO} in {LTE}-advanced: {SU-MIMO} vs. {MU-MIMO},'' \emph{{IEEE} Commun.
  Mag.}, vol.~50, no.~2, pp. 140--147, Feb. 2012.

\bibitem{Pados99}
I.~N. Psaromiligkos, S.~N. Batalama, and D.~A. Pados, ``On adaptive minimum
  probability of error linear filter receivers for {DS-CDMA} channels,''
  \emph{{IEEE} Trans. Commun.}, vol.~47, no.~7, pp. 1092--1102, Jul. 1999.

\bibitem{BlosteinCOM07}
N.~Wang and S.~D. Blostein, ``Approximate minimum {BER} power allocation for
  {MIMO} spatial multiplexing systems,'' \emph{{IEEE} Trans. Commun.}, vol.~55,
  no.~1, pp. 180--187, Jan. 2007.

\bibitem{BlosteinVT07}
------, ``Minimum {BER} transmit power allocation and beamforming for two-input
  multiple-output spatial multiplexing systems,'' \emph{{IEEE} Trans. Veh.
  Technol.}, vol.~56, no.~2, pp. 704--709, Mar. 2007.

\bibitem{Alouini13}
Q.~Z. Ahmed, M.-S. Alouini, and S.~Aissa, ``Bit error-rate minimizing detector
  for amplify-and-forward relaying systems using generalized {Gaussian}
  kernel,'' \emph{{IEEE} Signal Process. Lett.}, vol.~20, no.~1, pp. 55--58,
  Jan. 2013.

\bibitem{Barry97}
C.-C. Yeh and J.~R. Barry, ``Approximate minimum bit-error rate equalization
  for binary signaling,'' in \emph{Proc. {IEEE} Int. Conf. Commun. (ICC)},
  1997, pp. 1095--1099.

\bibitem{Barry00}
C.~C. Yeh and J.~R. Barry, ``Adaptive minimum bit-error rate equalization for
  binary signaling,'' \emph{{IEEE} Trans. Commun.}, vol.~48, no.~7, pp.
  1226--1235, Jul. 2000.

\bibitem{Antoniou00}
X.~Wang, W.-S. Lu, and A.~Antoniou, ``Constrained minimum-{BER} multiuser
  detection,'' \emph{{IEEE} Trans. Signal Process.}, vol.~48, no.~10, pp.
  2903--2909, Oct. 2000.

\bibitem{WeiAndChen07}
J.~Li, G.~Wei, and F.~Chen, ``On minimum-{BER} linear multiuser detection for
  {DS-CDMA} channels,'' \emph{{IEEE} Trans. Signal Process.}, vol.~55, no.~3,
  pp. 1093--1103, Mar. 2007.

\bibitem{Hanzo05}
S.~Chen, N.~N. Ahmad, and L.~Hanzo, ``Adaptive minimum bit error rate
  beamforming,'' \emph{{IEEE} Trans. Wireless Commun.}, vol.~4, no.~2, pp.
  341--348, Mar. 2005.

\bibitem{Hanzo08}
S.~Chen, A.~Livingstone, H.-Q. Du, and L.~Hanzo, ``Adaptive minimum symbol
  error rate beamforming assisted detection for quadrature amplitude
  modulation,'' \emph{{IEEE} Trans. Wireless Commun.}, vol.~7, no.~4, pp.
  1140--1145, Apr. 2008.

\bibitem{MajidQBSC14}
M.~Bavand, P.~Azmi, and S.~D. Blostein, ``Convex optimization based minimum
  probability of error beamforming in the uplink of a multiuser system,'' in
  \emph{Proc. {IEEE} 27th Biennial Symp. Commun. (QBSC)}, Jun. 2014, pp.
  28--32.

\bibitem{Schwarz16}
``Emerging communication technologies enabling the {I}nternet of things,''
  \emph{Rohde \& Schwarz White Paper}, Sep. 2016.

\bibitem{Nokia15}
``{LTE-M} - optimizing {LTE} for the {I}nternet of things,'' \emph{Nokia
  Network White Paper}, 2015.

\bibitem{Saeed14}
S.~Abdallah and S.~D. Blostein, ``Rate adaptation using long range channel
  prediction based on discrete prolate spheroidal sequences,'' in \emph{Proc.
  {IEEE} 15th Int. Workshop Signal. Process. Adv. Wireless Commun. (SPAWC)},
  Jun. 2014, pp. 479--483.

\bibitem{Forney98}
G.~D. Forney~Jr. and G.~Ungerboeck, ``Modulation and coding for linear
  {Gaussian} channels,'' \emph{{IEEE} Trans. Inf. Theory}, vol.~44, no.~6, pp.
  2384--2415, Oct. 1998.

\bibitem{PaulrajBook03}
A.~Paulraj, R.~Nabar, and D.~Gore, \emph{Introduction to Space-Time Wireless
  Communications}.\hskip 1em plus 0.5em minus 0.4em\relax Cambridge University
  Press, 2003.

\bibitem{Mao12}
J.~Mao, J.~Gao, Y.~Liu, and G.~Xie, ``Simplified semi-orthogonal user selection
  for {MU-MIMO} systems with {ZFBF},'' \emph{{IEEE} Wireless Commun. Lett.},
  vol.~1, no.~1, pp. 42--45, Feb. 2012.

\bibitem{NocedalBook06}
J.~Nocedal and S.~J. Wright, \emph{Numerical Optimization}, 2nd~ed.\hskip 1em
  plus 0.5em minus 0.4em\relax New York, USA: Springer, 2006.

\bibitem{BoydBook04}
S.~Boyd, \emph{Convex Optimization}.\hskip 1em plus 0.5em minus 0.4em\relax
  Cambridge University Press, 2004.

\bibitem{Brandwood83}
D.~Brandwood, ``A complex gradient operator and its application in adaptive
  array theory,'' \emph{Proc. {IEEE}}, vol. 130, no.~1, pp. 11--16, Feb. 1983.

\bibitem{GesbertTSP07}
A.~Hj{\o}rungnes and D.~Gesbert, ``Complex-valued matrix differentiation:
  Techniques and key results,'' \emph{{IEEE} Trans. Signal Process.}, vol.~55,
  no.~6, pp. 2740--2746, Jun. 2007.

\bibitem{Koivunen10}
J.~Eriksson, E.~Ollila, and V.~Koivunen, ``Essential statistics and tools for
  complex random variables,'' \emph{{IEEE} Trans. Signal Process.}, vol.~58,
  no.~10, pp. 5400--5408, Oct. 2010.

\bibitem{Peel05}
C.~B. Peel, B.~M. Hochwald, and A.~L. Swindlehurst, ``A vector-perturbation
  technique for near-capacity multiantenna multiuser communication-part {I}:
  Channel inversion and regularization,'' \emph{{IEEE} Trans. Commun.},
  vol.~53, no.~1, pp. 195--202, Jan. 2005.

\bibitem{Sadek07WC}
M.~Sadek, A.~Tarighat, and A.~H. Sayed, ``A leakage-based precoding scheme for
  downlink multi-user {MIMO} channels,'' \emph{{IEEE} Trans. Wireless Commun.},
  vol.~6, no.~5, pp. 1711--1721, May 2007.

\end{thebibliography}

\end{document}